\documentclass[aps,pre,reprint,groupedaddress,showpacs]{revtex4-1}

\usepackage{amsmath,amssymb}
\usepackage{graphicx}
\usepackage{subfigure}
\usepackage{bm}

\newcommand{\nn}{\nonumber}
\newcommand{\ds}{\displaystyle}

\newtheorem{pro}{Proposition}
\newtheorem{con}{Conjecture}

\begin{document}

% Use the \preprint command to place your local institutional report
% number in the upper righthand corner of the title page in preprint mode.
% Multiple \preprint commands are allowed.
% Use the 'preprintnumbers' class option to override journal defaults
% to display numbers if necessary
%\preprint{}

%Title of paper
\title{On stable laws in a one-dimensional intermittent map\\equipped with a uniform invariant measure}

% repeat the \author .. \affiliation  etc. as needed
% \email, \thanks, \homepage, \altaffiliation all apply to the current
% author. Explanatory text should go in the []'s, actual e-mail
% address or url should go in the {}'s for \email and \homepage.
% Please use the appropriate macro foreach each type of information

% \affiliation command applies to all authors since the last
% \affiliation command. The \affiliation command should follow the
% other information
% \affiliation can be followed by \email, \homepage, \thanks as well.
\author{Soya Shinkai}
%\email{soya@aoni.waseda.jp}
\thanks{Present address: Department of Mathematical and Life Science, Graduate School of Science, Hiroshima University, Kagamiyama, Higashi-Hiroshima 739-8526, Japan.}
\author{Yoji Aizawa}
%\homepage[]{Your web page}
%\thanks{}
%\altaffiliation{}
\affiliation{Department of Applied Physics, Faculty of Science and Engineering, Waseda University, 3-4-1 Okubo, Tokyo, 169-8555 Japan}

%Collaboration name if desired (requires use of superscriptaddress
%option in \documentclass). \noaffiliation is required (may also be
%used with the \author command).
%\collaboration can be followed by \email, \homepage, \thanks as well.
%\collaboration{}
%\noaffiliation

%\date{\today}

\begin{abstract}
We investigate ergodic-theoretical quantities and large deviation properties of one-dimensional intermittent maps,
that have not only an indifferent fixed point but also a singular structure such that the uniform measure is invariant under the mapping.
The probability density of the residence time and the correlation function are found to behave polynomially: $f(m) \sim m^{-(\kappa+1)}$ and $C(\tau) \sim \tau^{-(\kappa-1)}$ $(\kappa > 1)$.
Using the Doeblin-Feller theorems in probability theory, we derive the conjecture that the rescaled fluctuations of the time average of some observable functions obey the stable distribution with the exponent $1 < \alpha \le 2$.
Some exponents of the stable distribution are precisely determined by numerical simulations, and the conjecture is verified numerically.
The polynomial decay of large deviations is also discussed, and it is found that the entropy function does not exist, because the moment generating function of the stable distribution can not be defined.
\end{abstract}

% insert suggested PACS numbers in braces on next line
% insert suggested keywords - APS authors don't need to do this
%\keywords{}

%\maketitle must follow title, authors, abstract, \pacs, and \keywords
\maketitle

% body of paper here - Use proper section commands
% References should be done using the \cite, \ref, and \label commands
%\section{}
% Put \label in argument of \section for cross-referencing
%\section{\label{}}
%\subsection{}
%\subsubsection{}
%%%%%%%%%%%%%%%%%%%%  SECTION 1  %%%%%%%%%%%%%%%%%%%%%%%%%%%%%%%%%%%%%%%%
\section{Introduction}
It is well known that the chaotic motions of Hamiltonian dynamical systems exhibit slow dynamics, such as the polynomial decay of the time correlation function and the $\omega^{-\nu}$ power spectrum~\cite{Aizawa89b, Karney83,ChirikovShepelyansky84,GeiselZacherlRadons87}.
These slow motions should be understood from the measure-theoretical viewpoint.
However, the mixed phase space, which consists of integrable (torus) and non-integrable (chaos) components, makes it difficult to study the ergodic properties, despite the fact that the Liouville measure is invariant under the Hamiltonian flow.
Stagnant motions near the invariant tori add to the difficulty because of the appearance of non-stationary processes such as Arnold diffusion.
The stagnant layer theory based on the Nekhoroshev theorem has succeeded in explaining the scaling laws~\cite{Aizawa89a}, the induction phenomena in the lattice vibration~\cite{Aizawa89b} and the appearance of the log-Weibull distribution in $N$-body systems~\cite{AizawaSatoIto00}.

Intermittency is another area that typically exhibits slow dynamics.
Recently, the theory of intermittency has been developed using infinite ergodic theory, where the indifferent fixed points play an essential role in inducing the infinite measure and generating the non-stationary processes~\cite{Aaronson97,Aizawa00}.
The Darling-Kac-Aaronson theorem states that the rescaled time average of an $L^1_+$ observable function behaves essentially as the Mittag-Leffler (ML) random variable in the non-stationary dynamical processes;
the Lempel-Ziv complexity, which is a quantity of symbolic data in information theory, is also the ML random variable~\cite{ShinkaiAizawa06}.
Furthermore, for some classes of observable functions, the rescaled time average behaves as a random variable that is related to the ML random variable~\cite{Akimoto08}.
If there exists stationary processes in the theory of intermittency, the time average converges to the phase average, as the Birkhoff ergodic theorem describes.
However, the central limit theorem, which is usually regarded as a distributional law in stationary processes, can be violated:
anomalous fluctuations around the mean value and slow distributional convergence can appear.
Thus far, using renewal theory, large deviation theory and the semi-Markovian approximation, these anomalous behaviors have been understood~\cite{Aizawa89c,KikuchiAizawa90,TanakaAizawa93}.

In 2009, the polynomial decay of the large deviations for some observables was proved~\cite{Melbourne09,PollicottSharp09} and also shown numerically~\cite{ArtusoManchein09} in slowly mixing dynamical systems.
However, the distributional law has not yet been explained completely from the viewpoint of ergodic theory.
One of our objectives is to obtain numerical results of the distributional law of such anomalous fluctuations in order to develop the ergodic theory of the slow dynamics in the future.

This paper is organized as follows.
In Section~2, we introduce a class of one-dimensional intermittent maps equipped with a uniform invariant measure, motivated by the Pikovsky map~\cite{Pikovsky91} and the Miyaguchi map~\cite{MiyaguchiAizawa07}, and we analyse the statistical aspects.
In Section~3, we introduce several theorems in probability theory.
We also suggest a conjecture for the distributional law of the time averages of some observable functions from the viewpoint of ergodic theory.
In Section~4, we present our numerical results.
Finally, Section~5 is devoted to a summary and discussion.

%%%%%%%%%%%%%%%%%%%%  SECTION 2  %%%%%%%%%%%%%%%%%%%%%%%%%%%%%%%%%%%%%%%%
\section{\label{s2}The model and its statistical features}
In the last few decades, it has been recognized that one-dimensional intermittent maps are simple models for discussing the anomalous transport properties in chaotic Hamiltonian dynamical systems~\cite{GeiselThomae84,KlagesRadonsSokolv08}.
In these studies, the Pikovsky map is the first model equipped with a uniform invariant measure~\cite{Pikovsky91}.
Thermodynamical formalism and anomalous transport for the map have been studied using the periodic orbit theory~\cite{ArtusoCristadoro04}.
However, since the map is defined implicitly and is inadequate for extended numerical simulations, its ergodic properties have not yet been studied.
Miyaguchi and Aizawa improved the Pikovsky map via piecewise-linearisation and defining it explicitly, and they investigated its spectral properties using the Frobenius-Perron operator.

In this section, we introduce a map by which the Miyaguchi map is smoothly transformed and show that the uniform measure is approximately invariant under this map.
Furthermore, we analyse the statistical aspects and symmetrise the map in order to provide for the formalization and numerical simulations in the following sections.

%%%%%%%%%%%%%%%%%%%%  SUB SECTION 2.1  %%%%%%%%%%%%%%%%%%%%%%%%%%%%%%%%%%%%%%%%
\subsection{The map equipped with a uniform invariant measure}\label{ss2.1}
\begin{figure}
	\centering
	\subfigure[]{
		\includegraphics[width=0.9\hsize]{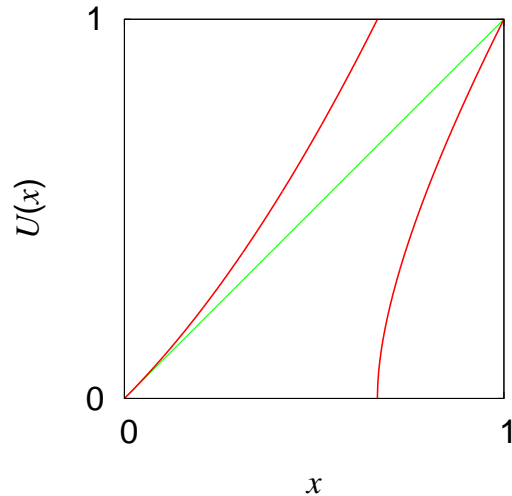}
		\label{mapU}	
	}
	\subfigure[]{
		\includegraphics[width=0.9\hsize]{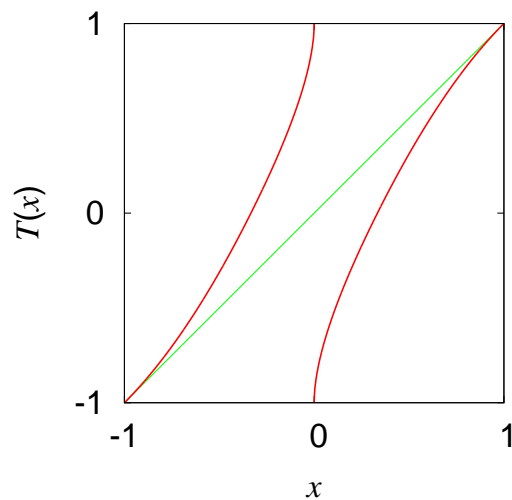}
		\label{mapT}
	}
	\caption{(Color online) (a) A graph of the map $U(x)$ and (b) a graph of the map $T(x)$ for $g(t) = t^2$ and $a = 2/3$.}
\end{figure}

Here we consider the map $U$ defined in the interval $[0,1]$ as follows:
\begin{equation}\label{def_map_U}
	U(x) = \left\{
	\begin{array}{lcl}
		\ds U_0(x) = x + (1-a) \, g\left( \frac{x}{a} \right) & \mbox{for} & x \in [0,a),\\
		\ds U_1(x) = x - a + a \, g^{-1}\left( \frac{x-a}{1-a} \right) & \mbox{for} & x \in [a,1],
	\end{array}
	\right.
\end{equation}
where $a$ is a constant ($0<a<1$) and function $g: [0,1] \to [0,1]$ has the following properties:
\begin{itemize}
\item
	the inverse function $g^{-1}$ exists;
\item
	$g(0) = 0$ and $g(1) = 1$;
\item
	for $t \ll 1$, $g(t) \ll t$ $\left( g^{-1}(t) \gg t \right)$ and $g'(t) \ll 1$ $\left( \left( g^{-1} \right)'(t) \gg 1 \right)$.
\end{itemize}
This map, where $g(t) = t^2$ and $a = 2/3$, is shown in Fig.~\ref{mapU}, and the left part $U_0(x)$ is the same as the Pomeau-Manneville-type intermittent maps.
The right part $U_1(x)$, however, is different from such intermittent maps, and the derivative $U'_1(x)$ is divergent at $x = a$ for $g^{-1}(t) = t^{1/2}$. 

Given these properties for $g$, the uniform density can be approximately derived as a solution of the Frobenius-Perron equation of the map $U$ as follows:
Let $b \, (< a)$ be a some small constant.
For the points $y_0 \in [0,a)$ and $y_1 \in [a,1]$ that satisfy $x = U(y_i) \, (i=0,1)$,
if $y_0 /a \ll 1$ and $(y_1-a)/(1-a) \ll 1$ for small $x < b$,
we can write approximately
\begin{eqnarray}
	\rho(x) & = & \sum_{y_i \in U^{-1}x} \frac{\rho(y_i)}{U'(y_i)} \nn\\
	 & = & \rho(y_0) \left\{ 1 + \frac{1-a}{a} g'\left( \frac{y_0}{a} \right) \right\}^{-1} \nn\\
	 & & + \rho(y_1) \left\{ 1 + \frac{a}{1-a} \left(g^{-1}\right)' \left( \frac{y_1-a}{1-a} \right) \right\}^{-1} \nn\\
	 & \simeq & \rho(y_0) \left\{ 1 - \frac{1-a}{a} g'\left( \frac{y_0}{a} \right) \right\} \nn \\
	 & & + \rho(y_1) \left\{ \frac{a}{1-a} \left(g^{-1}\right)' \left( \frac{y_1-a}{1-a} \right) \right\}^{-1}, \label{approFP}
\end{eqnarray}
where $\rho$ is an invariant density of the map $U$.
We can also obtain the approximate relations
\begin{equation*}
	x \simeq y_0 \quad \mbox{and} \quad g(x/a) \simeq \frac{y_1-a}{1-a}.
\end{equation*}
Substituting these expressions and the formula for differentiation of an inverse function
\begin{equation*}
	\left( g^{-1} \right)' \left( g(t) \right) = \left\{ g'(t) \right\}^{-1}
\end{equation*}
into Eq.~(\ref{approFP}), we obtain for small $x < b$
\begin{equation}\label{FPsmallx}
	\rho(x) \simeq \rho(y_0) \left\{ 1 - \frac{1-a}{a} g' \left( \frac{x}{a} \right) \right\}
	+ \rho(y_1) \frac{1-a}{a} g' \left( \frac{x}{a} \right).
\end{equation}
For $x>b$, we assume that the slopes of $U_0$ and $U_1$ can be approximately equal to $\ds \frac{1-b}{a-b}$ and $\ds \frac{1-b}{1-a}$, respectively.
Then the Frobenius-Perron equation can be approximately written as
\begin{equation}\label{FPfinitex}
	\rho(x) \simeq \rho(y_0) \, \frac{a-b}{1-b} + \rho(y_1) \, \frac{1-a}{1-b}.
\end{equation}
Therefore, Eqs.~(\ref{FPsmallx}) and (\ref{FPfinitex}) imply that the uniform density $\rho(x)=1$ is an approximate solution of the Frobenius-Perron equation of the map $U$ for $x \in (0,1)$.

%%%%%%%%%%%%%%%%%%%%  SUB SECTION 2.2  %%%%%%%%%%%%%%%%%%%%%%%%%%%%%%%%%%%%%%%%
\subsection{Residence time distribution}\label{ss2.2}
Here, under the mapping $U$, we consider the residence time $m$, which is a random variable,
of an orbit in the interval $[0,a)$ and then
asymptotically estimate the probability density function $f(m)$.

\begin{figure}
	\centering
	\includegraphics[width=\hsize]{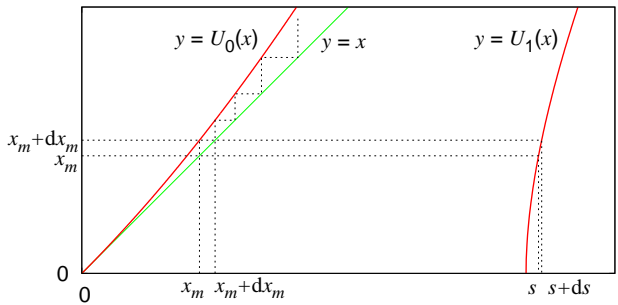}
	\caption{(Color online) Points $x_m$ and $s$ satisfy $U_0^m(x_m) = 1$ and $U_1(s)=x_m$.
	The image $U_1\left( [s,s+ds) \right)$ corresponds to the interval $[x_m,x_m+dx_m)$.}
	\label{map_U0_U1}
\end{figure}

Let us consider the points $x_m \in [0,a)$ and $s \in [a,1]$ such as $U_0^m(x_m) = 1$ and $U_1(s) = x_m$ (see Fig.~\ref{map_U0_U1}).
First, we estimate the relationship between $x_m$ and $m$ as follows:
For $\Delta x_m \equiv U_0(x_m) - x_m$ and $\Delta m \equiv m - (m-1)$,
the left part of Eq.~(\ref{def_map_U}) can be written as
\begin{equation}\label{differenceEq}
	\frac{\Delta x_m}{\Delta m} = (1-a) g(x_m/a).
\end{equation}
Using the continuous approximation for $x_m \ll 1 \, (m \gg 1)$ , Eq.~(\ref{differenceEq}) becomes
\begin{equation}\label{dx/dm}
	\frac{d x}{d m} \simeq (1-a) g(x/a).
\end{equation}
Now let us integrate Eq.~(\ref{dx/dm}) as follows:
\begin{eqnarray}\label{mG}
	m = \int_0^m \, d m' & \simeq & \frac{1}{1-a} \int_{x_m}^a \frac{d x}{g(x/a)}\nn\\
	& = & \frac{a}{1-a} \int_{x_m/a}^1 \frac{d y}{g(y)} \quad (y=x/a)\nn\\
	& = & \frac{a}{1-a} \left\{ G(x_m/a) - G(1) \right\},
\end{eqnarray}
where the function $G(x)$ is the indefinite integral $-\int^x d y / g(y)$.
Let us assume that $G(x_m/a) \gg G(1)$ for $x_m/a \ll 1$ and that the function $G$ has an inverse $G^{-1}$.
Then Eq.~(\ref{mG}) can be approximated as $m \sim G(x_m/a)$ and we can obtain the relationship
\begin{equation}\label{x_m-m}
	x_m / a \sim G^{-1}(m).
\end{equation}

Next, we consider the probability of the injection orbits into the interval $[x_m,x_m+d x_m)$.
Under the continuous approximation,
we assume that the interval $[s,s+d s)$ is mapped onto the interval $[x_m,x_m+d x_m)$, as shown in Fig.~\ref{map_U0_U1}.
The orbit that starts from the interval $[x_m,x_m+d x_m)$ resides for $m$ time-steps in the interval $[0,a).$
Therefore, using the invariant measure $\rho$, the residence time probability density is defined as
\begin{equation}\label{res_measure}
	f(m) dm \equiv \rho(s) ds.
\end{equation}
From the definition of point $s$ and under the condition $\ds g^{-1}\left( \frac{s-a}{1-a} \right) \gg \frac{s-a}{1-a}$,
we can estimate
\begin{equation*}
	g(x_m/a) \simeq \frac{s-a}{1-a}.
\end{equation*}
Using Eqs.~(\ref{dx/dm}), (\ref{x_m-m}) and (\ref{res_measure}) and the approximation $\rho(s) \simeq 1$,
we can estimate the residence time probability density as
\begin{eqnarray}\label{estimation_res_dis}
	f(m) & \sim & g'(x_m/a) \frac{d x_m}{d m}\nn\\
	& \sim & g'\left(G^{-1}(m)\right) \cdot g\left( G^{-1}(m) \right) \quad (m \gg 1).
\end{eqnarray}

%%%%%%%%%%%%%%%%%%%%  SUB SECTION 2.3  %%%%%%%%%%%%%%%%%%%%%%%%%%%%%%%%%%%%%%%%
\subsection{The symmetrised map and its statistical features}\label{ss2.3}

%%%%%%%%%%%%%%%%%%%%  SUB SUB SECTION 2.3.1  %%%%%%%%%%%%%%%%%%%%%%%%%%%%%%%%%%%%%%%%
\subsubsection{The model}
Let us symmetrise the map $U$ into map $T$ defined on the interval $[-1,1]$ as follows:
\begin{widetext}
\begin{equation*}
	T(x) = \left\{
	\begin{array}{lclcl}
		U_0(x+1) -1 & = &
		\ds x + (1-a) \, g\left( \frac{x+1}{a} \right) & \mbox{for} & x \in [-1,-1+a),\\
		-U_1(-x+a) + 1 & = &
		\ds x + 1 - a \, g^{-1}\left( \frac{-x}{1-a} \right) & \mbox{for} & x \in [-1+a,0),\\
		U_1(x+a) - 1 & = &
		\ds x - 1 + a \, g^{-1}\left( \frac{x}{1-a} \right) & \mbox{for} & x \in [0,1-a),\\
		-U_0(-x+1) + 1 & = &
		\ds x - (1-a) \, g\left( \frac{-x+1}{a} \right) & \mbox{for} & x \in [1-a,1].
	\end{array}
	\right.
\end{equation*}
\end{widetext}
In the following, we also assume that the function $g$ is given by
\begin{equation}\label{gt}
	g(t) = t^B,
\end{equation}
where $B \ge 1$ is a parameter and the properties in Subsection~\ref{ss2.1} are satisfied, and assume that
\begin{equation*}
	a = B / (B+1),
\end{equation*}
so that the derivative $T'(x)$ is continuous at $x=-1+a$ and $1-a$.
Fig.~\ref{mapT} shows a graph of the map $T$ for $B=2$.

%%%%%%%%%%%%%%%%%%%%  SUB SUB SECTION 2.3.2  %%%%%%%%%%%%%%%%%%%%%%%%%%%%%%%%%%%%%%%%
\subsubsection{Residence time probability density}
The analytical estimates for the map $U$ in Subsections~\ref{ss2.1} and \ref{ss2.2} are essentially the same as for the map $T$.
Therefore, using Eqs.~(\ref{estimation_res_dis}) and (\ref{gt}), we can derive the probability density of the residence time in each interval $[-1,0)$ and $[0,1]$ as
\begin{equation}\label{res_time_dis}
	f(m) \sim m^{-(\kappa+1)} \quad ( m \gg 1, \, \kappa = B / (B-1) ).
\end{equation}
Note that $B=2 \, (\kappa = 2)$ is a transition point with the second moment $\langle m^2 \rangle \equiv \int_1^\infty m^2 f(m) \, d m$ being finite for $1 \le B < 2 \, (\kappa > 2)$ but infinite for $B > 2 \, (1 < \kappa < 2)$.
Comparing this with our previous results for the modified Bernoulli map, we can classify the parameter region as follows~\cite{Aizawa89c,KikuchiAizawa90}:
\begin{enumerate}
\item
	$1 \le B < 2$ $(\kappa > 2)$, Gaussian stationary region,
\item
	$B > 2$ $(1 < \kappa < 2)$, Non-Gaussian stationary region.
\end{enumerate}

%%%%%%%%%%%%%%%%%%%%  SUB SUB SECTION 2.3.3  %%%%%%%%%%%%%%%%%%%%%%%%%%%%%%%%%%%%%%%%
\subsubsection{Correlation function}
By transforming orbits $\{ T^n(x) \}$ into coarse-grained orbits $\left\{ \sigma \left( T^n(x) \right) \right\}$, where the function $\sigma$ is defined by
\begin{equation*}
	\sigma(x) = \left\{
	\begin{array}{lcl}
		-1 & \mbox{for} & x < 0,\\
		+1 & \mbox{for} & x > 0,
	\end{array}
	\right.
\end{equation*}
we can calculate the correlation function $C(\tau)$ by renewal analysis~\cite{Aizawa84}:
\begin{equation}\label{cor_fun}
	C(\tau) \sim \tau^{-(\kappa - 1)} \quad (\tau \gg 1).
\end{equation}
Given that the quantity $\int C(\tau) \, d \tau$ is a time-scale criterion of the random fluctuations~\footnote{When $C(\tau)$ decays exponentially, the quantity corresponds to the relaxation time.}, we see that it is finite for $\kappa > 2$ but infinite for $1 < \kappa < 2$.
Consequently, for $B > 2 \, (1 < \kappa < 2)$, a remarkably long time tail is revealed.

%%%%%%%%%%%%%%%%%%%%  SECTION 3  %%%%%%%%%%%%%%%%%%%%%%%%%%%%%%%%%%%%%%%%
\section{Distributions of partial sums}\label{s3}
Here we state two probability theorems proved by Doeblin and Feller~\cite{Feller49,Feller71} and
suggest a conjecture for the distribution of the partial sums under the mapping $T$.
In part, the derivation of the conjecture is the same, as in our previous papers~\cite{Aizawa89c,KikuchiAizawa90,TanakaAizawa93}.

%%%%%%%%%%%%%%%%%%%%  SUB SECTION 3.1  %%%%%%%%%%%%%%%%%%%%%%%%%%%%%%%%%%%%%%%%
\subsection{Doeblin-Feller theorems}
\begin{figure}[t]
	\centering
	\includegraphics[width=0.9\hsize]{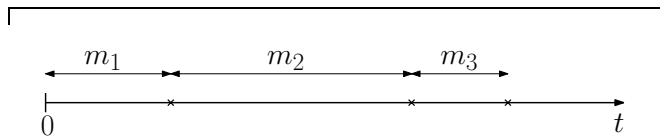}
	\caption{An event occurs at each cross.
	A sequence $m_j$ stands for the interval time between $j$th and $(j-1)$th events.}
	\label{recurrent_event}
\end{figure}

In probability theory, recurrent events in which the successive waiting times are mutually independent random variables have been the object of study (see Fig.~\ref{recurrent_event}).
One of the problems is as follows:
What distribution does the epoch of the $r$th occurrence, which is a random variable, obey?
And, how does it converge?
Assuming that the first and the second moments for successive waiting times exist,
the answer is given by the central limit theorem, which states that the rescaled fluctuations obey the Gauss distribution.
For the case where the first and/or the second moments do not exist, the answer was given by Doeblin and Feller.
Here we consider only the case in which the second moment does not exist.

\begin{pro}[Doeblin-Feller~\cite{Feller49}]\label{Pro1}
Let us consider a sequence of mutually independent positive random variables $\{ \bm{m}_j \}$
with a common distribution $F$, which satisfies the following properties:
\begin{equation*}
	1 - F(x) = x^{-\alpha} h(x),
\end{equation*}
where the function $h$ varies slowly at $\infty$, i.e., for every constant $s$
\begin{equation*}
	\lim_{x \to \infty} \frac{h(sx)}{h(x)} = 1.
\end{equation*}
Let $\bm{M}_r = \bm{m}_1 + \cdots + \bm{m}_r$ and define $b_r$ by
\begin{equation*}
	1 - F(b_r) \sim \frac{1}{r},
\end{equation*}
and denote by $\bm{N}_t$ the number of renewal events in the time interval $(0,t)$, i.e.,
\begin{equation*}
	\mathrm{Pr} \, \{ \bm{N}_t \ge r \} = \mathrm{Pr} \, \{ \bm{M}_r \le t \}.
\end{equation*}
Then, for $1 < \alpha < 2$ and $\mu = E(\bm{m}_j) < \infty$,
\begin{widetext}
\begin{eqnarray*}
	\mathrm{Pr} \, \left\{ \left( \frac{\bm{M}_r}{r} - \mu \right) r^{1-1/\alpha} \le x \right\}  & \to &
	V\left( x;\alpha,2-\alpha,c\right) \quad (r \to \infty),\\
	\mathrm{Pr} \, \left\{\mu^{1+1/\alpha}\left(\frac{\bm{N}_t}{t}-\frac{1}{\mu}\right)t^{1-1/\alpha} \le x \right\} & \to &
	V\left( x;\alpha,\alpha-2,c \right) \quad (t \to \infty),
\end{eqnarray*}
\end{widetext}
where $V$ denotes the stable distribution (see Appendix~\ref{AppA}) and $c=\left( -\Gamma(1-\alpha) \right)^{1/\alpha}$.
\end{pro}
This proposition states that the rescaled fluctuations for the random variables $\bm{M}_r$ and $\bm{N}_t$ obey the stable distribution, where the skewness parameters are different.
Note that for $\alpha > 2$ the limit distribution corresponds to the Gauss distribution $V(x;2,0,c)$ because the first and second moments for $F$ exist.

For real-valued random variables, the following proposition is implied by Feller and can be proven in the same way as the proof of Proposition 1.

\begin{pro}[Feller~\cite{Feller71}]\label{Pro2}
Let $\{ \bm{X}_j \}$ be a sequence of mutually independent real-valued random variables with a common distribution $\tilde{F}$ that satisfies the following properties:
\begin{equation*}
	1-\tilde{F}(x) = A_+ x^{-\alpha} h(x), \quad \tilde{F}(-x) = A_- x^{-\alpha} h(x) \quad (x>0),
\end{equation*}
\begin{equation*}
	\tilde{F}(0) = A_-, \quad A_+ \ge 0, \quad A_- \ge 0, \quad A_+ + A_- = 1.
\end{equation*}
Let $\bm{Y}_n = \bm{X}_1 + \cdots + \bm{X}_n$ and define $\tilde{b}_n$ by
\begin{equation*}
	1 - \tilde{F}(\tilde{b}_n) \sim \frac{A_+}{n}, \quad \tilde{F}(-\tilde{b}_n) \sim \frac{A_-}{n}.
\end{equation*}
Let $\tilde{\mu} = E(\bm{X}_j)$ be the mean value and
set the probability $Q_n(x) = \mathrm{Pr} \, \left\{ \left( \bm{Y}_n / n - \tilde{\mu} \right) n^{1-1/\alpha} \le x \right\}$.
Then, for $1 < \alpha < 2$,
\begin{widetext}
\begin{equation*}
	\int_{-\infty}^\infty e^{i z x} \, d Q_n(x) \to \exp \left[ - |z|^\alpha \left\{ \cos \frac{\pi\alpha}{2} \mp
	i (A_+ - A_-) \sin \frac{\pi\alpha}{2} \right\} \Gamma(1-\alpha) \right] \qquad (n \to \infty).
\end{equation*}
\end{widetext}
The expression on the right-hand side corresponds to the characteristic function of the stable distribution.
In particular, we have for the following special cases:
\begin{equation*}
	Q_n(x) \to \left\{
	\begin{array}{lcc}
		V(x;\alpha,\alpha-2,c) & \mbox{for} & A_+=0, \, A_-=1,\\
		V(x;\alpha,2-\alpha,c) & \mbox{for} & A_+=1, \, A_-=0,\\
		V(x;\alpha,0,c) & \mbox{for} & A_+=\frac{1}{2}, \, A_-=\frac{1}{2},	
	\end{array}
	\right.
\end{equation*}
as $n \to \infty$, where $c$ is a suitable scale parameter for each case.
\end{pro}
Proposition 2 states that the distribution $\tilde{F}$ belongs to the domain of attraction of the stable distribution $V$.

%%%%%%%%%%%%%%%%%%%%  SUB SECTION 3.2  %%%%%%%%%%%%%%%%%%%%%%%%%%%%%%%%%%%%%%%%
\subsection{Distributions of the partial sums for some observable functions under the mapping $T$}
\begin{figure*}
	\centering
	\subfigure[]{
		\includegraphics[scale=0.5]{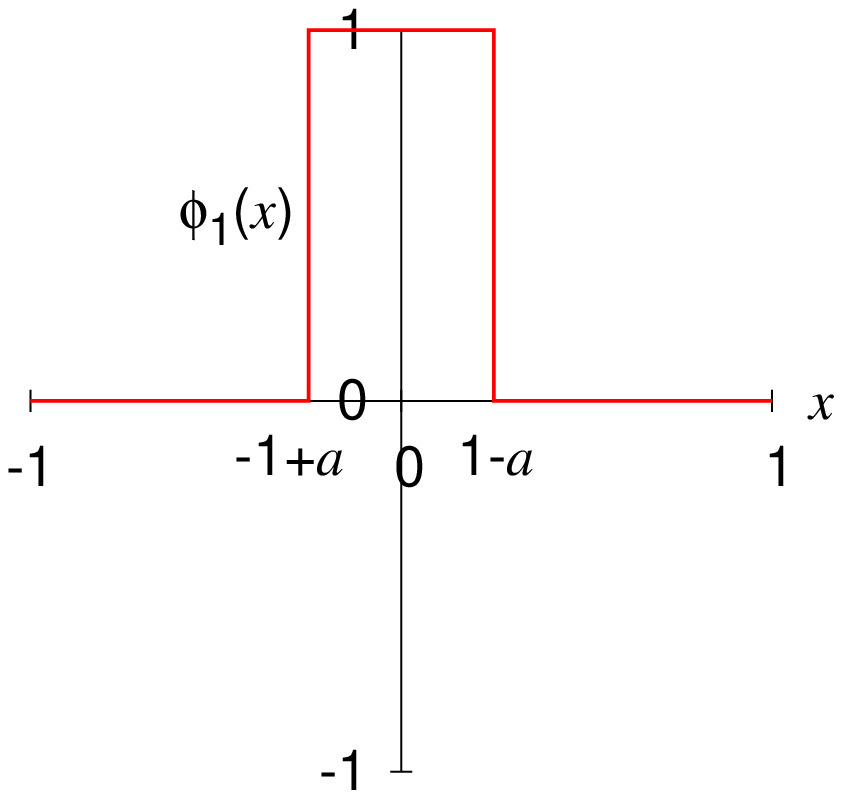}
		\label{phi1}	
	}
	\quad
	\subfigure[]{
		\includegraphics[scale=0.5]{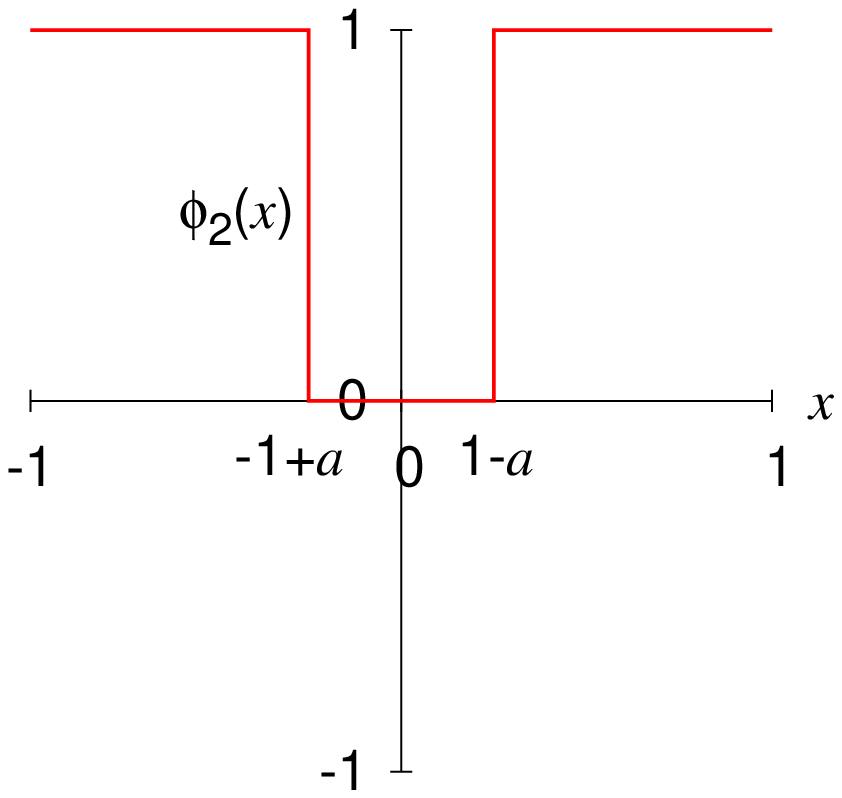}
		\label{phi2}
	}
	\quad
	\subfigure[]{
		\includegraphics[scale=0.5]{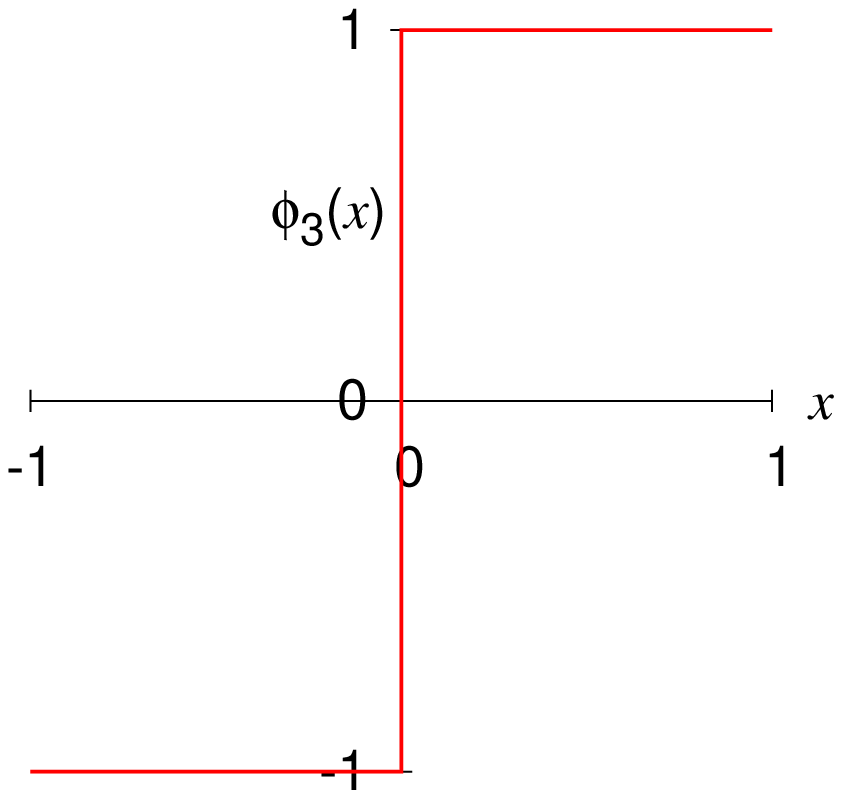}
		\label{phi3}
	}
	\caption{(Color online) The observable functions (a) $\phi_1$, (b) $\phi_2$ and (c) $\phi_3$.}
\end{figure*}
Here we consider partial sums under the mapping $T$
\begin{equation*}
	S_n(\phi_j) \equiv \sum_{k=0}^{n-1} \phi_j \circ T^k,
\end{equation*}
where the observable functions $\phi_j \, (j=1,2,3)$ are defined by
\begin{subequations}
\begin{eqnarray}
	\phi_1(x) & = & \left\{
	\begin{array}{ccl}
	+1 & \mbox{for} & |x| < 1-a,\\
	0 & \mbox{for} & \mbox{otherwise},
	\end{array}
	\right. \label{defphi1}\\
	\phi_2(x) & = & -\phi_1(x)+1, \label{defphi2}\\
	\phi_3(x) & = & \sigma(x). \label{defphi3}
\end{eqnarray}
\end{subequations}
These functions are shown in Figs.~\ref{phi1}-\ref{phi3}.
As shown in Subsection~\ref{ss2.3}, the map $T$ induces statistical features.
As a result, for sufficiently large $n$, we can regard the partial sums as random variables as follows:
\begin{subequations}
	\begin{eqnarray}
		S_n(\phi_1) & \simeq & \bm{N}_n, \label{Sn1}\\
		S_n(\phi_2) & \simeq & n - \bm{N}_n, \label{Sn2}\\
		S_n(\phi_3) & \simeq & \sum_{i=1}^{\bm{N}_n/2} (\bm{m}_i^+ - \bm{m}_i^-), \label{Sn3}
	\end{eqnarray}
\end{subequations}
where $\bm{N}_n$ denotes the number of times an orbit changes its sign in the time interval $(0,n)$ and
$\bm{m}_i^\pm$ is the $i$th resident time in the region $x \gtrless 0$ for an orbit.

Then, for $\phi_1$ and $\phi_2$, Proposition 1 can be applied.
For $\phi_3$, the distribution of the random variables $\{ \bm{m}_i^+ \}$ and $\{ -\bm{m}_i^- \}$ corresponds to the distribution $\tilde{F}$ in Proposition 2 for $A_+ = 1/2$ and $A_- = 1/2$.
Therefore, we can derive the following conjecture:
\begin{con}
	Under the mapping $T$ and for each observable function $\phi_j \, (j=1,2,3)$,
	let us assume that
	\begin{equation*}
		\bm{\xi}_{n,j} = \left(\frac{S_n(\phi_j)}{n} - \langle \phi_j \rangle \right) n^{1-1/\alpha}
	\end{equation*}
	is a random variable, where $\langle \cdot \rangle$ denotes the ensemble average.
	Then,
	\begin{equation}\label{Probability_of_xi}
	\mathrm{Pr} \, \left\{ \bm{\xi}_{n,j} \le x \right\} \to V(x;\alpha,\gamma_j,c_j) \quad \mbox{as} \quad n \to \infty,
	\end{equation}
	where $\alpha$ is the function of $B$
	\begin{equation}\label{theory_alpha}
		\alpha = \left\{
		\begin{array}{ccl}
			2 & \mbox{for} & B < 2,\\
			B / (B-1) & \mbox{for} & B > 2,
		\end{array}
		\right.
	\end{equation}
	$\gamma_j$ is different for each function
	\begin{equation}\label{gamma_j}
		\gamma_j = \left\{
		\begin{array}{ccl}
			\alpha - 2 & \mbox{for} & j = 1,\\
			2 - \alpha & \mbox{for} & j = 2,\\
			0 & \mbox{for} & j = 3,\\
		\end{array}
		\right.		
	\end{equation}
	and $c_j$ is a suitable scale parameter.
\end{con}
The reason why the transition point $B=2$ is excluded in Eq.~(\ref{theory_alpha}) is that we need take into account a logarithmic correction such as the second moment $\langle m^2 \rangle = \int_1^\infty m^2 f(m) \, dm \sim \bigl[ \log m \bigr]_1^\infty$ for $B=2$.
In the following Subsection~\ref{ss4.2} we are not concerned with the point $B=2$.

%%%%%%%%%%%%%%%%%%%%  SECTION 4  %%%%%%%%%%%%%%%%%%%%%%%%%%%%%%%%%%%%%%%%
\section{Numerical results}\label{s4}
%%%%%%%%%%%%%%%%%%%%  SUB SECTION 4.1  %%%%%%%%%%%%%%%%%%%%%%%%%%%%%%%%%%%%%%%%
\subsection{Entropy, residence time distribution and correlation function}
\begin{figure}
	\centering
	\includegraphics[width=\hsize]{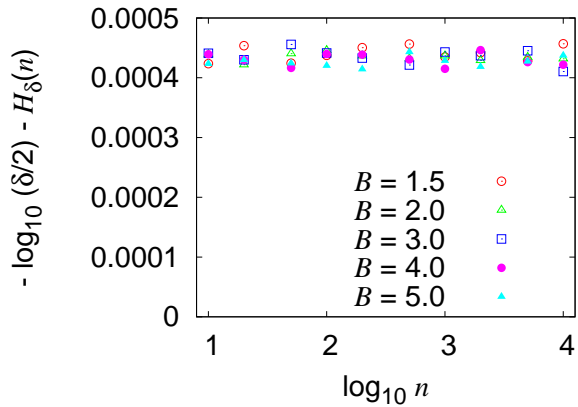}
	\caption{(Color online) Difference between $-\log_{10} (\delta/2)$ and equipartition entropies $H_\delta(n)$ for five different parameters $B$ under the mapping $T$ ($\delta = 10^{-3}$).
	For each $B$, $10^6$ initial points are uniformly distributed in the interval $(-1,1)$.}
	\label{entropy}
\end{figure}

First, in order to check that the uniform measure is invariant under the mapping $T$, we calculate the equipartition entropy
\begin{equation*}
	H_\delta (n) = -\sum_{i=-\delta^{-1}}^{\delta^{-1}-1} \lambda_n(D_i) \log_{10} \lambda_n(D_i),
\end{equation*}
where $\delta$ is the length of the set $D_i = (i \delta, (i+1)\delta) \, (i = -\delta^{-1}, -\delta^{-1}+1,\cdots,\delta^{-1}-1)$,
$\lambda_n(D_i)$ denotes the probability measure of the set $D_i$ at time-step $n$, and
an initial ensemble is uniformly given on the interval $(-1,1)$. 
In ergodic theory, the upper bound of the entropy is $-\log_{10}(\delta/2)$~\cite{ArnoldAvez68}.
For $\delta = 10^{-3}$ and five different parameters $B$,
the numerical results are shown in Fig.~\ref{entropy}, where $10^6$ initial points are distributed.
Because $-\log_{10} (\delta/2) - H_\delta(n)$ is less than $5 \times 10^{-4}$,
the uniform measure is numerically invariant under the mapping $T$.

\begin{figure*}
	\centering
	\subfigure[]{
		\includegraphics[width=0.5\hsize]{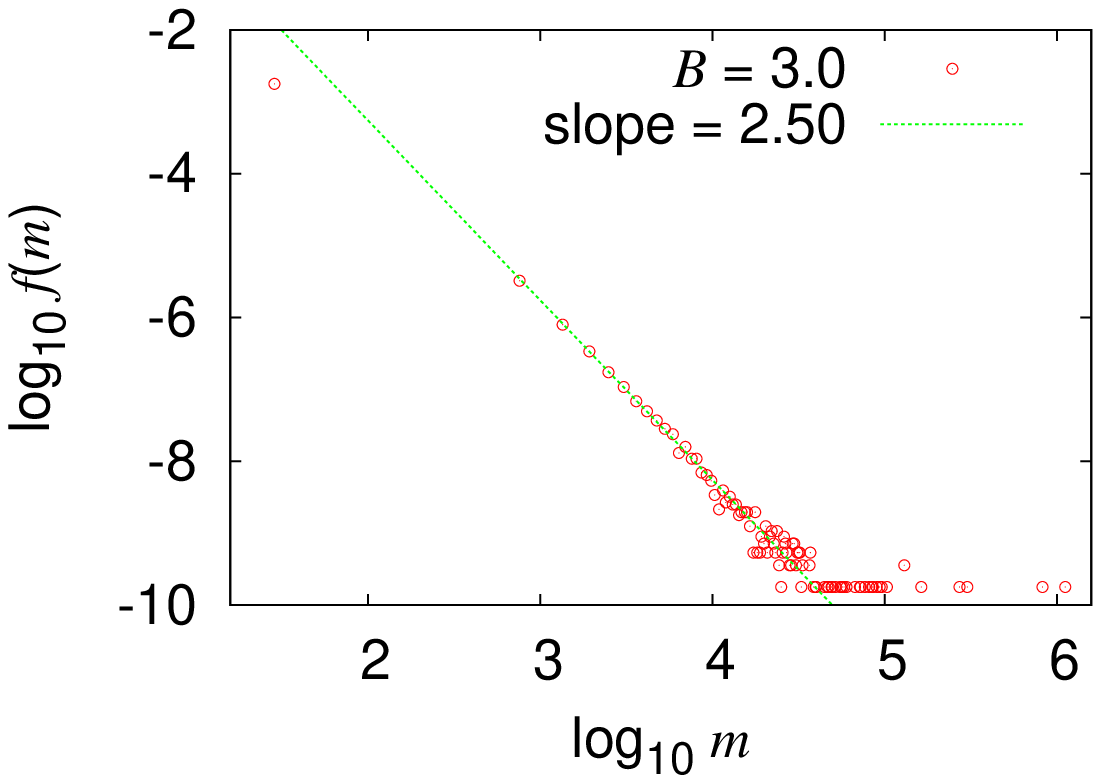}
		\label{log-log_res_time_dis}
	}
	\quad
	\subfigure[]{
		\includegraphics[width=0.45\hsize]{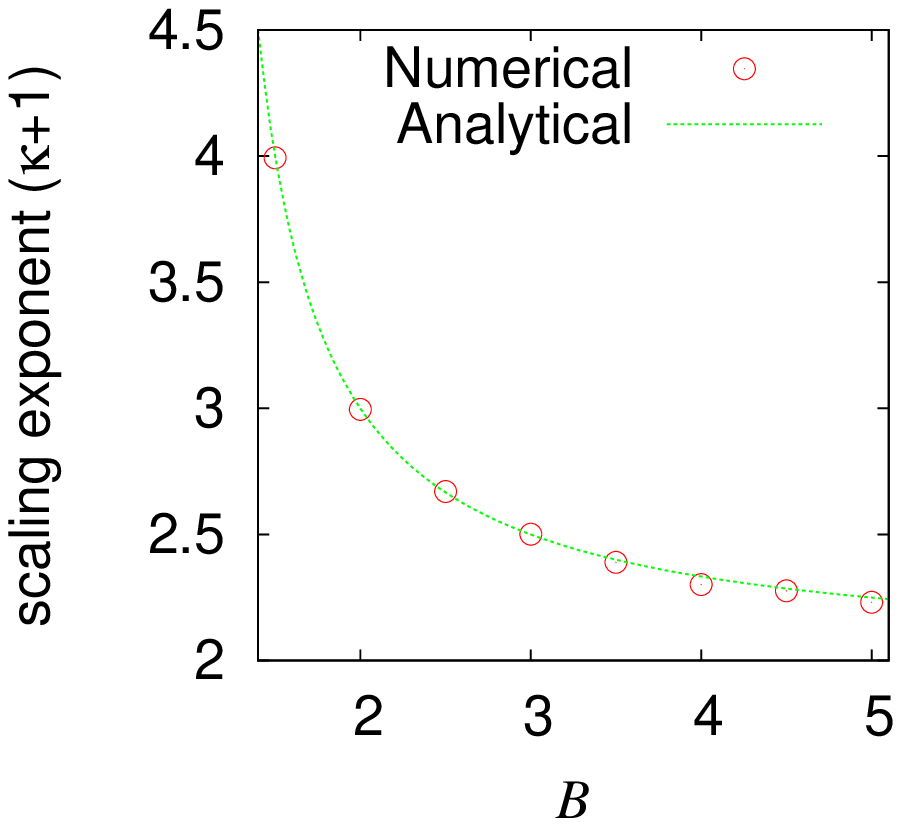}
		\label{scaling_exponent_res_time_dis}
	}
	\caption{(Color online) (a) The log-log plot of the residence time probability density $f(m)$.
	The circles represent the numerical result for $B=3.0$, and
	the dashed line is obtained by least-squares fitting in the large time-step ($10^3< m < 10^4$).
	(b) The scaling exponent of the probability density as a function of $B$.
	The circles represent the slopes by fitting as shown in (a), and the dashed curve is the theoretical result.}
\end{figure*}

Second, Fig.~\ref{log-log_res_time_dis} shows the probability density of the residence time $f(m)$ for $B=3.0$.
We can clearly see a power law $f(m) \sim m^{-2.5}$.
The scaling exponents for some parameters $B$ are shown in Fig.~\ref{scaling_exponent_res_time_dis}, and
the analytical result (see Eq.~(\ref{res_time_dis})) is verified.

\begin{figure*}
	\centering
	\subfigure[]{
		\includegraphics[width=0.5\hsize]{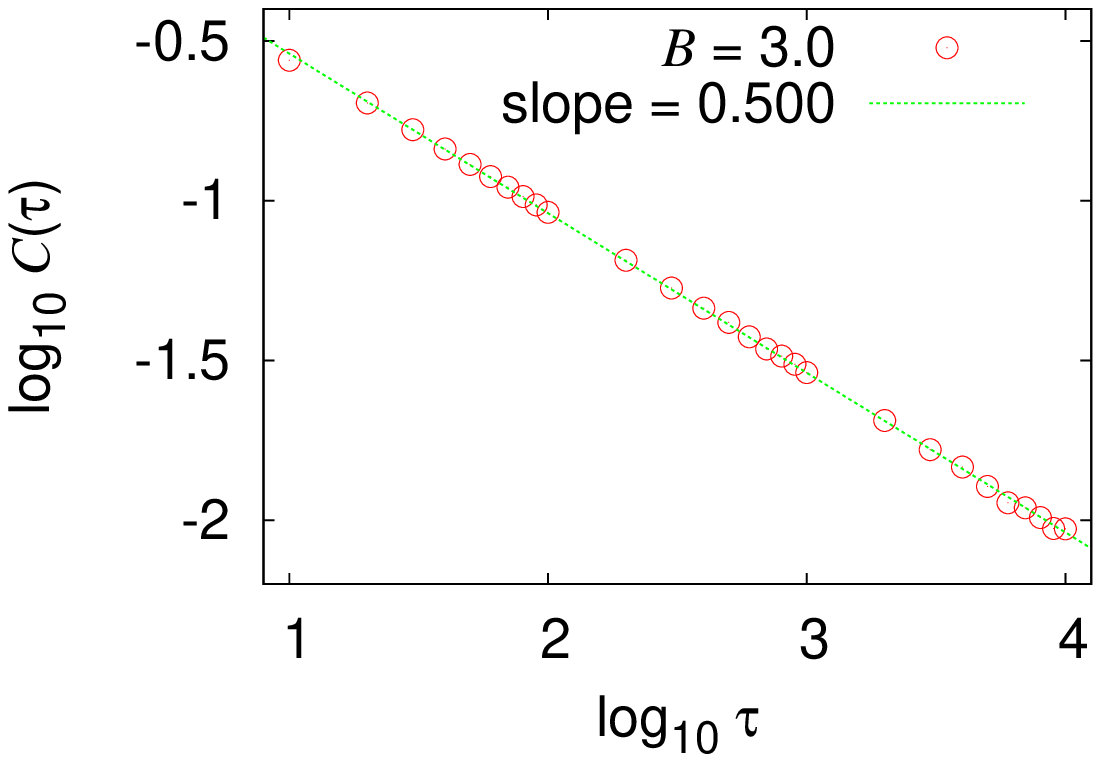}
		\label{log-log_cor_fun}
	}
	\quad
	\subfigure[]{
		\includegraphics[width=0.45\hsize]{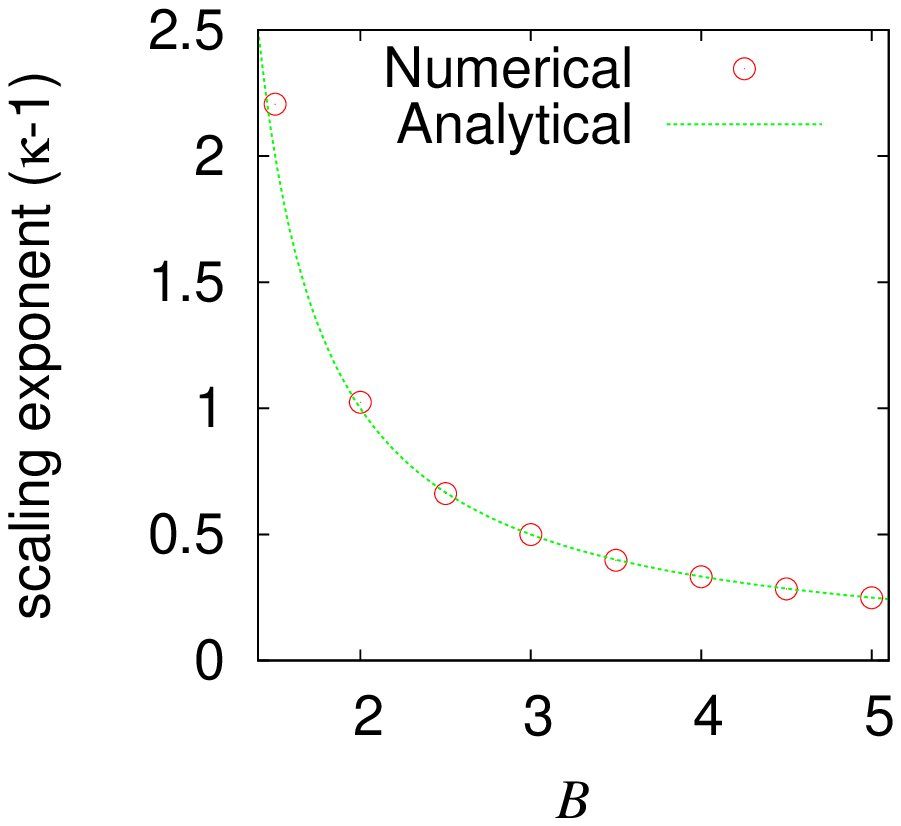}
		\label{scaling_exponent_cor_fun}
	}
	\caption{(Color online) (a) The log-log plot of the correlation function $C(\tau)$.
	The circles represent the numerical result for $B=3.0$, and
	the dashed line is obtained by least-squares fitting in the large time-step ($10^1 \le \tau \le 10^4$).
	(b) The scaling exponent of the correlation function $C(\tau) \sim \tau^{-(\kappa-1)}$ as a function of $B$.
	The circles represent the slopes by fitting as shown in (a), and the dashed curve is the theoretical result.}
\end{figure*}

Third, in numerical simulations the correlation function is calculated as
\begin{equation*}
	C(\tau) = \left| \left\langle \sigma(x) \sigma(T^\tau x) \right\rangle \right|,
\end{equation*}
where $\langle \cdot \rangle$ stands for the ensemble average and
$10^7$ initial points are given uniformly in the interval $(-1,1)$.
Figure~\ref{log-log_cor_fun} shows the correlation function for $B=3.0$.
%A power law $C(\tau) \sim \tau^{-0.5}$ is clearly observed.
The scaling exponents for some parameters $B$ are shown in Fig.~\ref{scaling_exponent_cor_fun}, and
the analytical result (see Eq.~(\ref{cor_fun})) is verified.

%%%%%%%%%%%%%%%%%%%%  SUB SECTION 4.2  %%%%%%%%%%%%%%%%%%%%%%%%%%%%%%%%%%%%%%%%
\subsection{Distributions of fluctuations of partial sums for some observable functions}\label{ss4.2}
\begin{figure*}
	\centering
	\subfigure[]{
		\includegraphics[width=0.47\hsize]{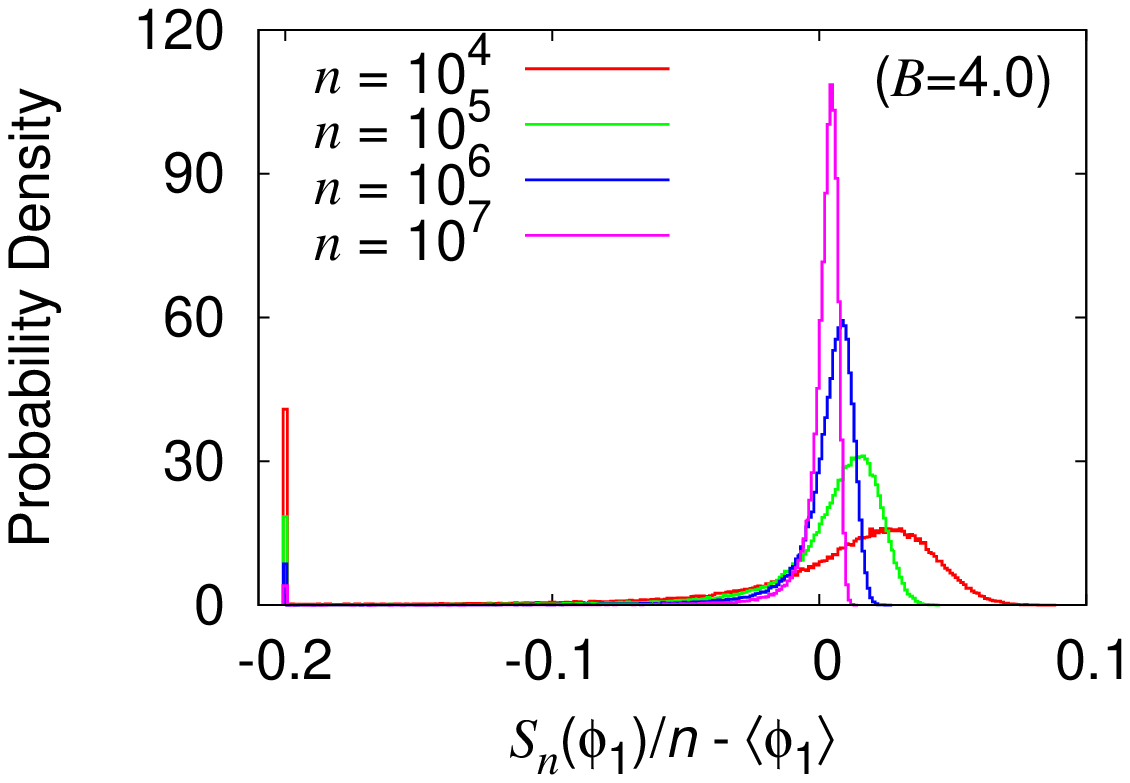}
		\label{pd_phi1}
	}
	\subfigure[]{
		\includegraphics[width=0.47\hsize]{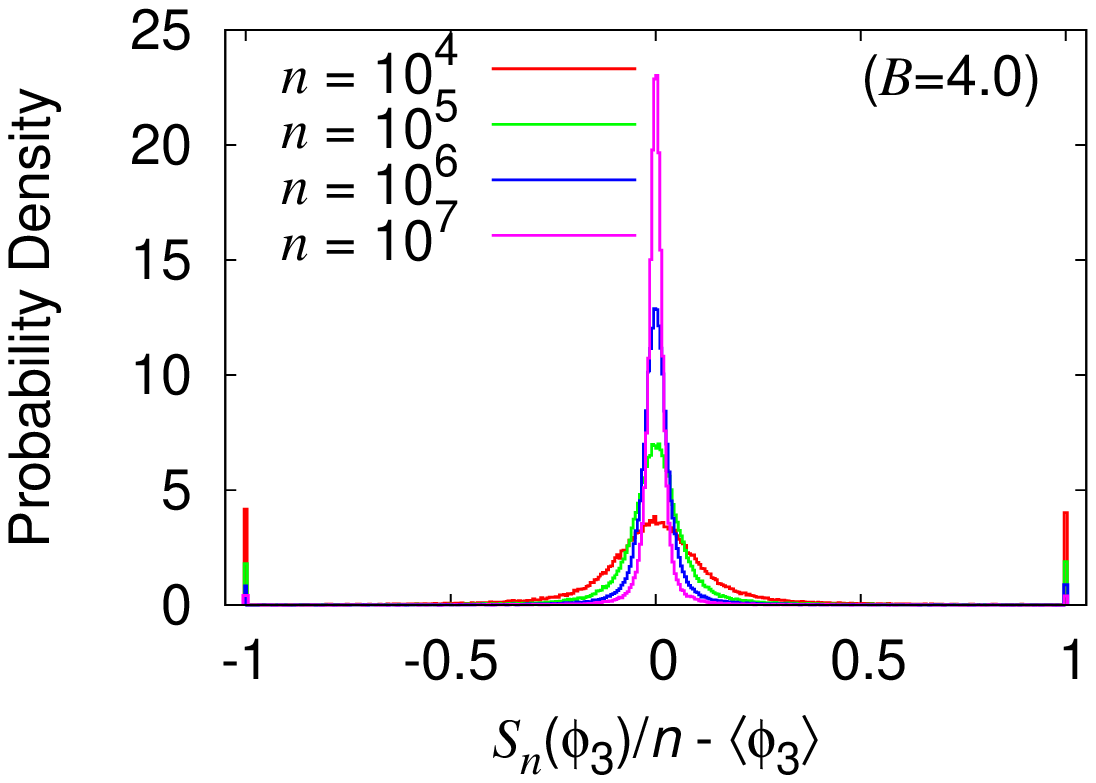}
		\label{pd_phi3}
	}
	\caption{(Color online) The probability densities of $S_n(\phi_j)/n - \langle \phi_j \rangle$ for the map $T$ with $B=4.0$ at $n=10^4,\,10^5,\,10^6$ and $10^7$.
	The indexes of the observable function $\phi_j$ are $j=1$ in (a) and $j=3$ in (b).}
\end{figure*}

Figures~\ref{pd_phi1} and \ref{pd_phi3} show the probability densities of $S_n(\phi_j)/n - \langle \phi_j \rangle$ ($j = 1, 3$) for $B = 4.0$ and $n = 10^4 \sim 10^7$.
In numerical simulations, $10^5$ initial points are provided in the interval $(-1,1)$.
As shown in our previous papers~\cite{KikuchiAizawa90,TanakaAizawa93} on the non-Gaussian stationary region $(1 < \alpha < 2)$,
the probability density consists of two components.
One is observed at $- \langle \phi_1 \rangle$ $(=-1/(B+1))$ in Fig.~\ref{pd_phi1} or $\pm 1$ in Fig.~\ref{pd_phi3} and decays when $n$ increases.
The other converges to the $\delta$-measure at the origin.
Note that the first component appears for $B \ge 2$ and relates to the distribution of the first passage time to escape around $x = \pm 1$,
but in what follows we neglect this component of the probability density.
Here we numerically analyse the normalized second component.

\begin{figure*}
	\centering
	\subfigure[]{
		\includegraphics[width=0.45\hsize]{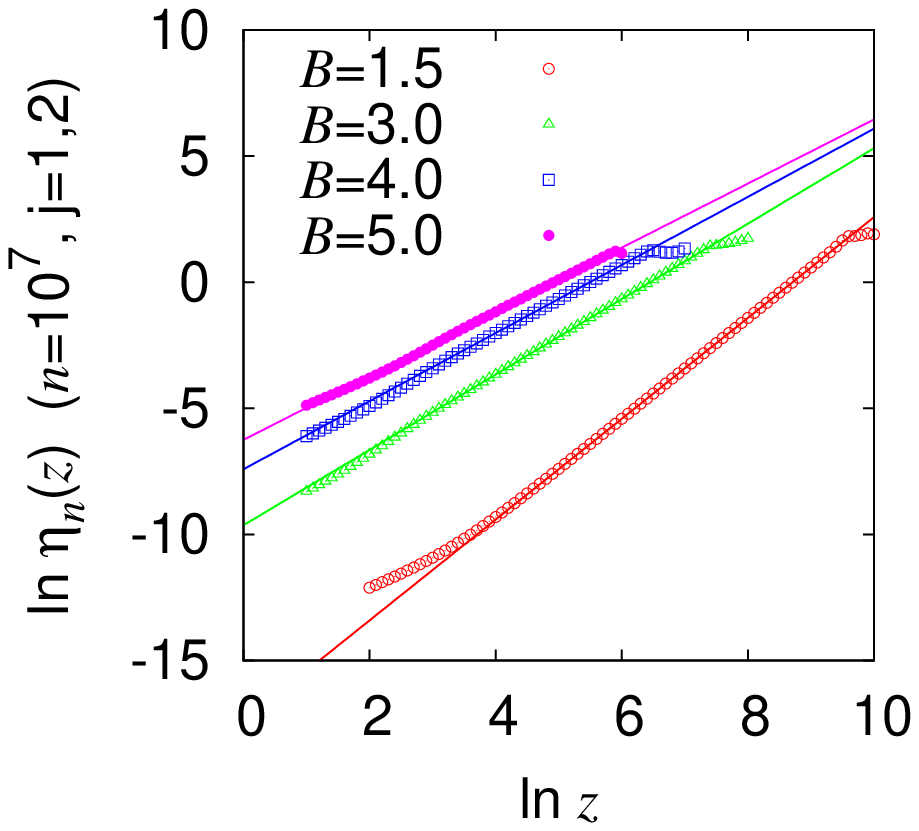}
		\label{fig9a}
	}
	\quad
	\subfigure[]{
		\includegraphics[width=0.45\hsize]{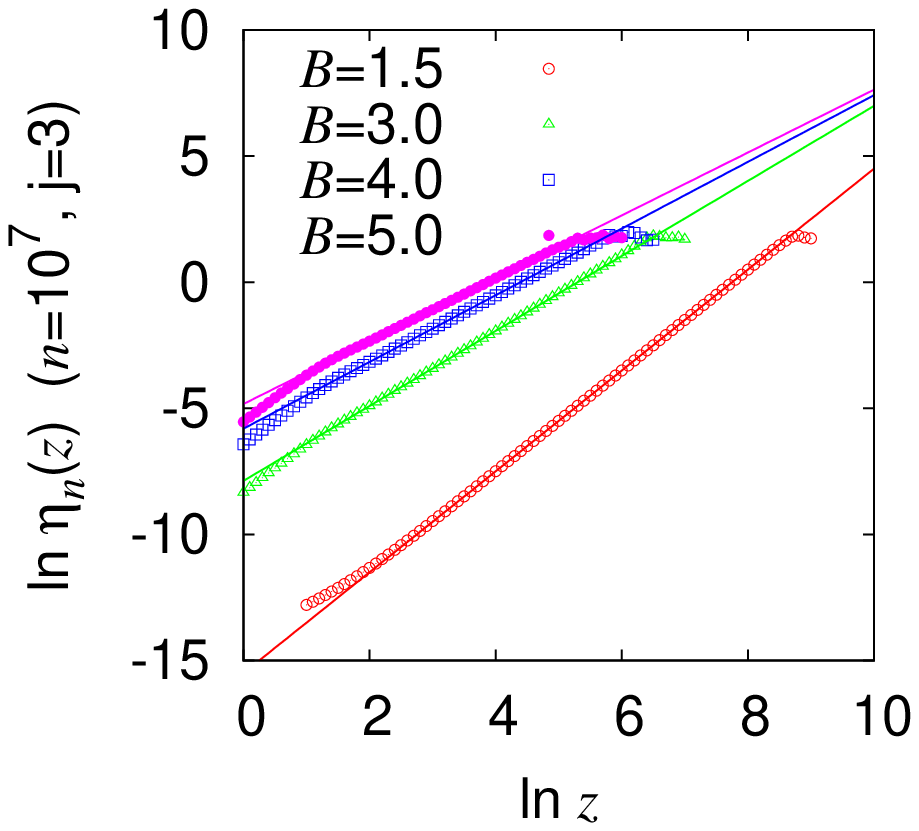}
		\label{fig9b}
	}
	\\
	\subfigure[]{
		\includegraphics[width=0.45\hsize]{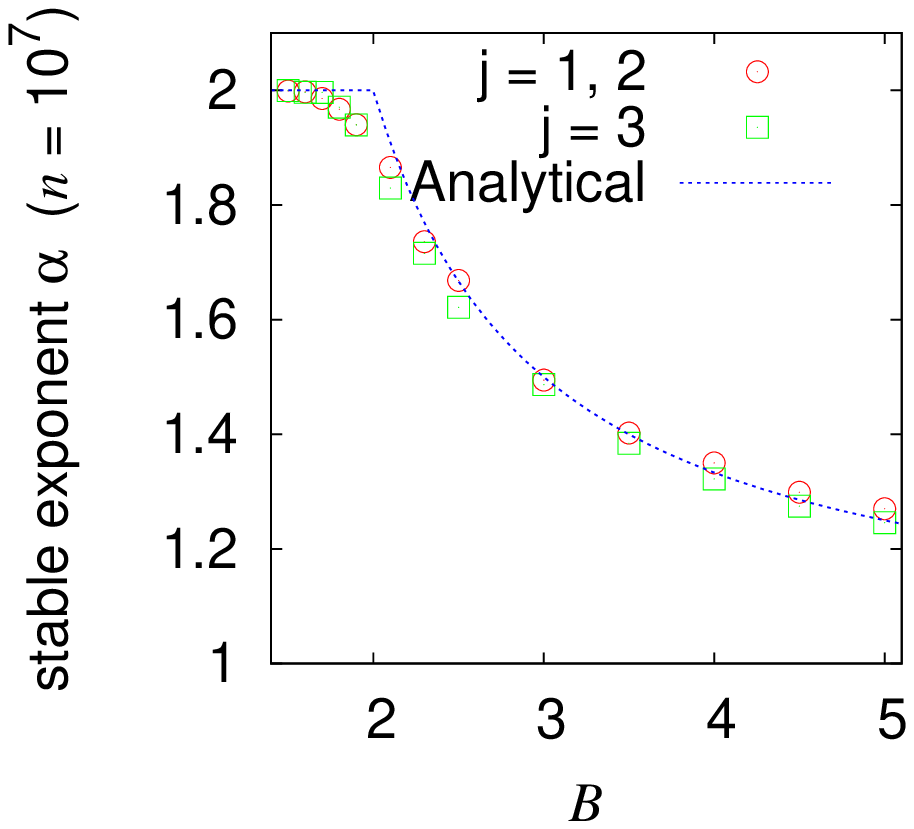}
		\label{fig9c}
	}
	\quad
	\subfigure[]{
		\includegraphics[width=0.45\hsize]{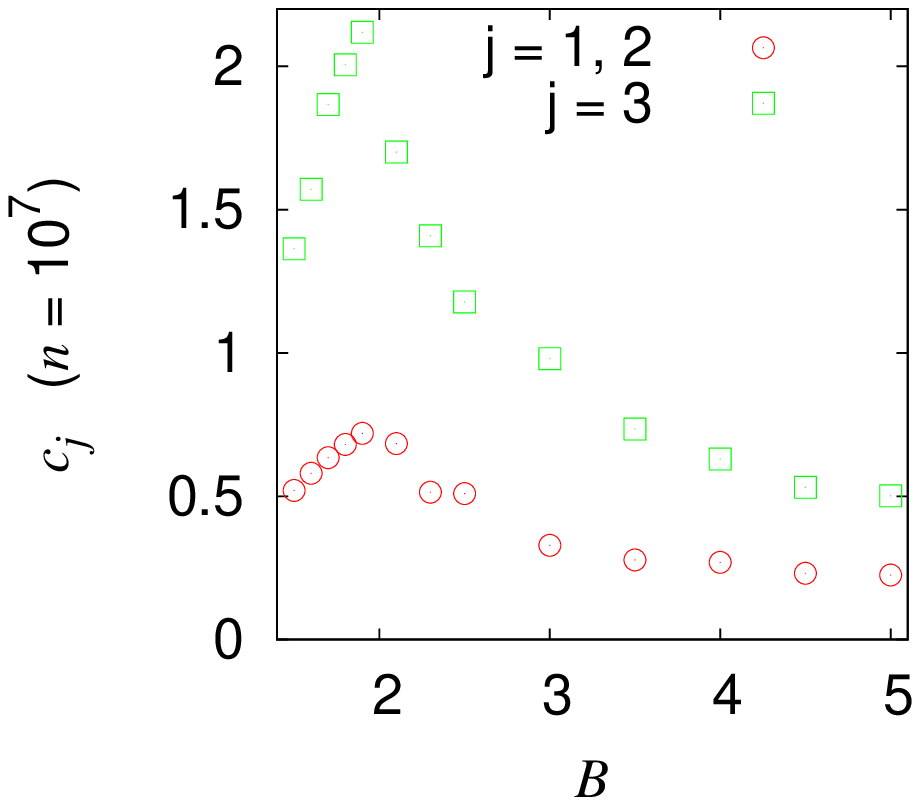}
		\label{fig9d}
	}
	\caption{(Color online) (a) and (b) stand for log-log plots of the function $\eta_n(z)$ (points), which is calculated from the characteristic function of the normalized probability density of $S_n(\phi_j)/n-\langle\phi_j\rangle$ at $n=10^7$ for $j=1,2$ and $j=3$, respectively.
	Lines are obtained by least-squares fitting by use of Eq.~(\ref{log_eta}).
	For $B = 1.5, \, 3.0,\ 4.0$ and $5.0$, numerically obtained values of $\alpha$ and $c_j$ are listed in Table~\ref{table_parameters}.
	(c) Numerically obtained stable exponent $\alpha$ as a function of $B$ for $j=1,\,2$ (circle) and $j=3$ (box).
	The dashed curve represents the analytical result as shown in Eq.~(\ref{theory_alpha}).
	(d) Numerically obtained scale parameters $c_j$ for $j=1,\,2$ (circle) and $j=3$ (box).
}
\end{figure*}

\begin{figure*}
	\centering
	\subfigure[]{
		\includegraphics[width=0.47\hsize]{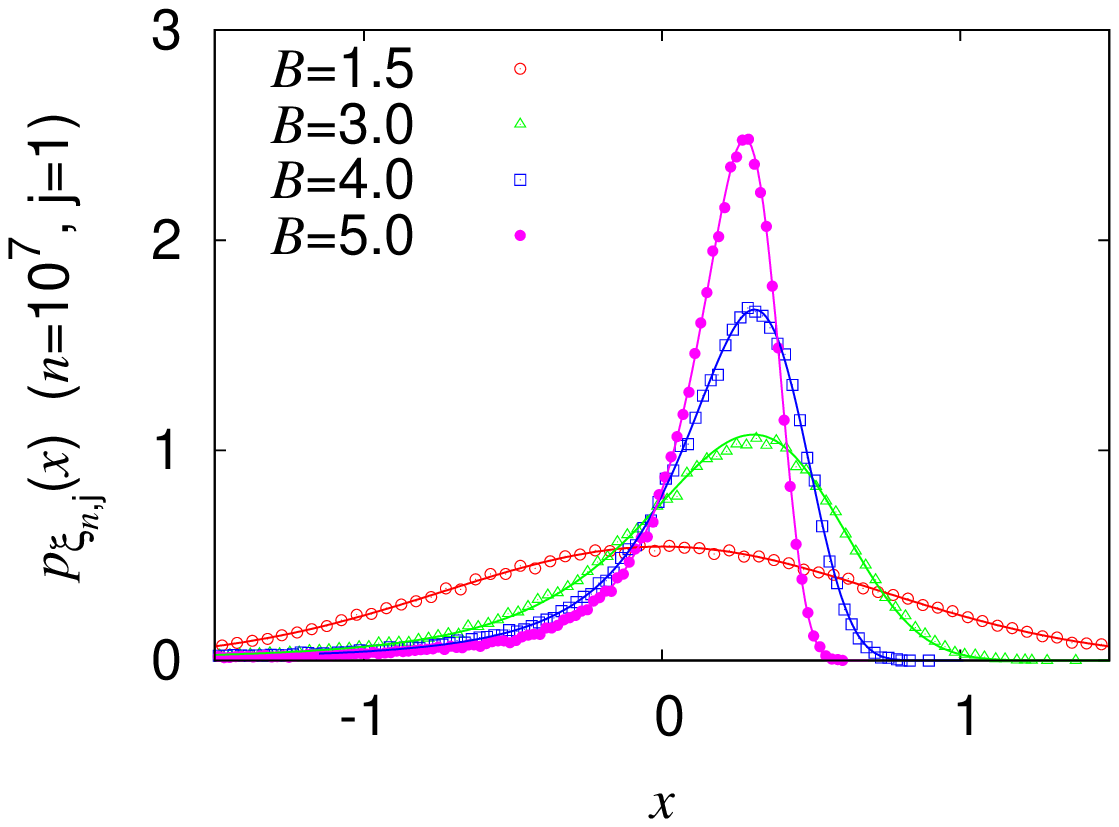}
		\label{fig10a}
	}
	\subfigure[]{
		\includegraphics[width=0.47\hsize]{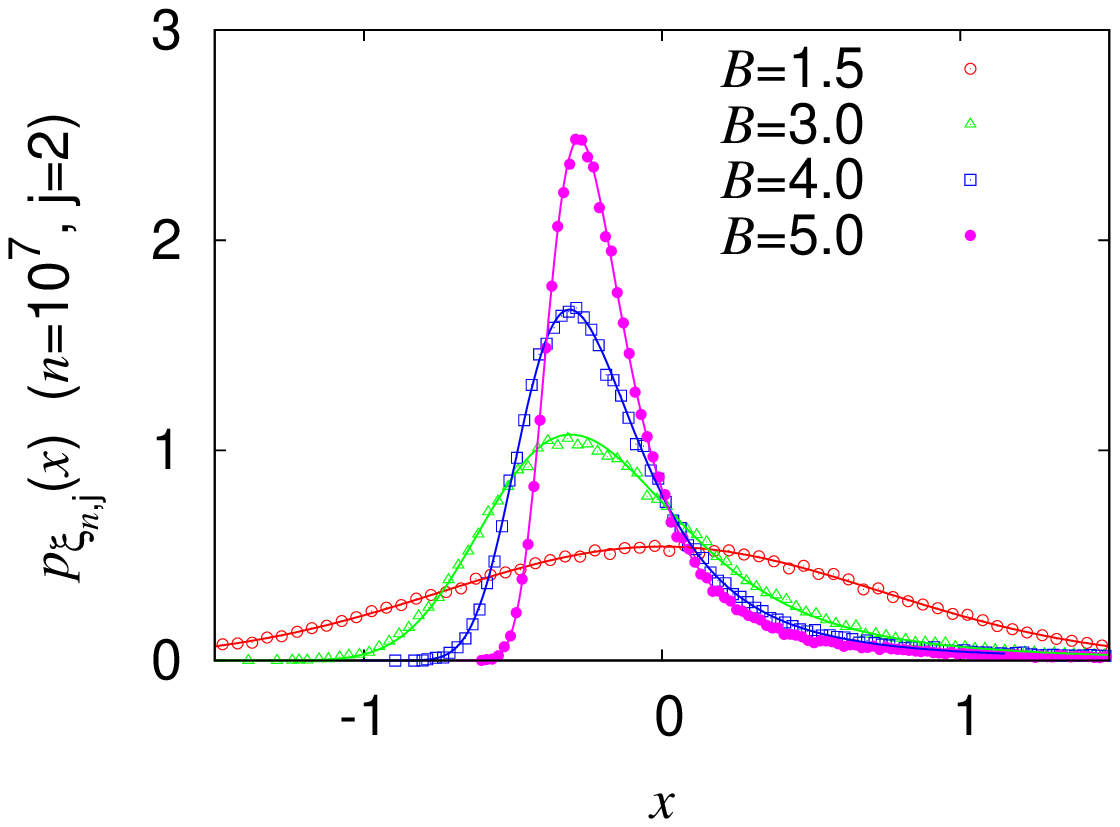}
		\label{fig10b}
	}
	\subfigure[]{
		\includegraphics[width=0.47\hsize]{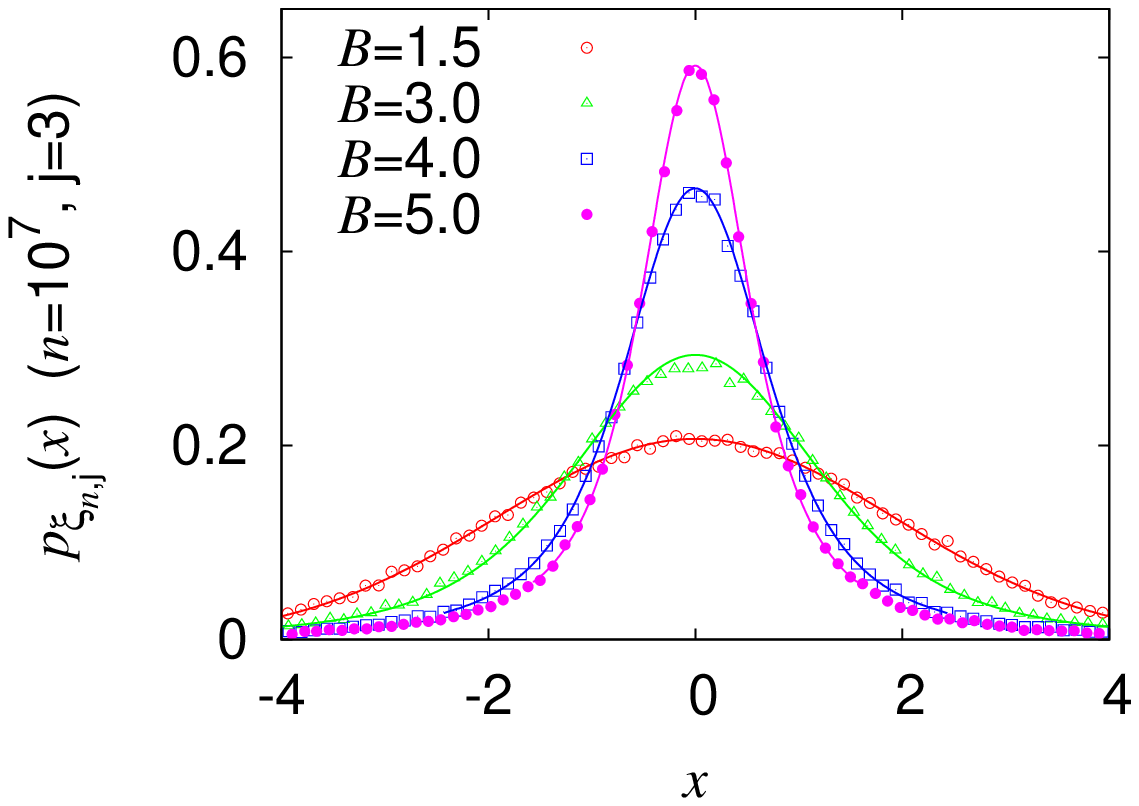}
		\label{fig10c}
	}
	\caption{(Color online) The probability densities $p_{\bm{\xi}_{n,j}}(x)$ (points) and the stable densities $v(x;\alpha,\gamma_j,c_j)$ (solid curves)
	for $B=1.5$ (circle), 3.0 (triangle), 4.0 (box) and 5.0 (filed circle) at $n=10^7$.
	The indexes of the observable functions $\phi_j$ are (a) $j=1$, (b) $j=2$ and (c) $j=3$.
	The values of $\alpha$ and $c_j$ of the stable density are shown in Table~\ref{table_parameters}.
	The values of the skewness parameter $\gamma_j$ are in Eq.~(\ref{gamma_j}).}
\end{figure*}

Given that the analysed density is the stable density,
we can estimate the stable exponent $\alpha$ and the scale parameter $c$ by use of the method as shown in Appendix~\ref{AppB}.
In our numerical simulations, the function $\eta_n(z)$ is calculated from the analysed density, the random variable of which is $S_n(\phi_j)/n-\langle\phi_j\rangle$.
Figures~\ref{fig9a} and \ref{fig9b} show log-log plots of $\eta_n(z)$ at $n=10^7$ for $j=1,\,2$ and $j=3$, respectively.
Using Eq.~(\ref{log_eta}) and least-squares fitting, $\alpha$ and $c_j$ are numerically obtained.
These values are shown in Figs.~\ref{fig9c} and \ref{fig9d}.

\begin{table}
	\caption{\label{table_parameters}
	Numerically obtained values of $\alpha$ and $c_j$ for $B=1.5$, 3.0, 4.0 and 5.0.
	The stable densities in Figs.~\ref{fig10a}-(c) are calculated using these values.}
	\begin{ruledtabular}
	\begin{tabular}{lllll}
		$B$ & $\alpha$ ($j=1,2$) & $c_1 \, (=c_2)$ & $\alpha$ ($j=3$) & $c_3$\\
		\colrule
		1.5 & 1.99873 & 0.520165 & 1.99986 & 1.36350\\
		3.0 & 1.49467 & 0.329168 & 1.48682 & 0.980109\\
		4.0 & 1.35012 & 0.269364 & 1.32224 & 0.629975\\
		5.0 & 1.27017 & 0.225277 & 1.24602 & 0.501533\\
	\end{tabular}
	\end{ruledtabular}
\end{table}

Next, using the numerically obtained exponent $\alpha$, we calculate the probability density, $p_{\bm{\xi}_{n,j}}(x)$, of the distribution $\mathrm{Pr} \, \{ \bm{\xi}_{n,j} \le x \}$.
In Figs.~\ref{fig10a}-\ref{fig10c}, the probability densities $p_{\bm{\xi}_{n,j}}(x)$ (points) with $\alpha$ and
the stable densities $v(x;\alpha,\gamma_j,c_j)$ (solid curves) with $\alpha$ and $c_j$ are shown for $B=1.5$, 3.0, 4.0 and 5.0.
The values of $\alpha$ and $c_j$ are shown in Table~\ref{table_parameters}, and
the values of the skewness parameter $\gamma_j$ are shown in Eq.~(\ref{gamma_j}).
These numerical results support our conjecture.

%%%%%%%%%%%%%%%%%%%%  SECTION 5  %%%%%%%%%%%%%%%%%%%%%%%%%%%%%%%%%%%%%%%%
\section{Summary and discussion}\label{s5}
In this study, we introduced a class of one-dimensional intermittent maps equipped with a uniform invariant measure and analysed their statistical features.
Working from the Doeblin-Feller theorems, we suggested the conjecture that the rescaled fluctuation for sums of some observable functions is the {\em stable} random variable.
Numerical results clearly indicate that the conjecture should be true.

Observable functions $\phi_j$ defined in Eqs.~(\ref{defphi1})-(\ref{defphi3}) are typical in {\em renewal processes}~\cite{GodrecheLuck01};
Equations~(\ref{Sn1})-(\ref{Sn3}) can be interpreted as follows:
$S_n(\phi_1)$ is the number of renewals,
$S_n(\phi_2)$ is the total occupation time around indifferent fixed point
and $S_n(\phi_3)$ is the mean of the process.
In this sense, the choice of observable functions is reasonable.
In {\em ergodic theory}, however, it is necessary to discuss both the class of observable functions and the limit distribution of sums for functions.
This kind of problem is one of open problems~\cite{Sinai08}.
Gou\"ezel proved a limit theorem of the observable function $\phi(x)=x^{-\beta}$ for the Bernoulli map $x \mapsto 2x$ (mod 1)~\cite{Gouezel08}.
Akimoto discussed a similar limit theorem for an infinite measure dynamical system~\cite{Akimoto08}.

Large deviations derived from our conjecture polynomially decay as follows:
Let us assume Eq.~(\ref{Probability_of_xi}), i.e.,
\begin{equation*}
	\mathrm{Pr}\, \left\{ \frac{S_n(\phi_j)}{n} - \langle \phi_j \rangle \le \frac{x}{n^{1-1/\alpha}} \right\} = V(x).
\end{equation*}
Then, using the property of the stable distribution (\ref{property_stable}), we can derive a polynomial decay of large deviations for $\varepsilon = x / n^{1-1/\alpha} > 0$:
\begin{eqnarray}\label{large_deviations}
	\mathrm{Pr} \, \left\{ \left| \frac{S_n(\phi_j)}{n}-\langle\phi_j\rangle \right| > \varepsilon \right\} & = &
	1 - V(n^{1-1/\alpha}\varepsilon) \nn\\
	& & + V(-n^{1-1/\alpha}\varepsilon), \nn \\
	& \sim & \left( n^{1-1/\alpha}\varepsilon \right)^{-\alpha}, \nn\\
	& \sim &  n^{1-\alpha}.
\end{eqnarray}
The same estimate has been obtained by other mathematicians \cite{Melbourne09,PollicottSharp09} and also shown numerically~\cite{ArtusoManchein09} in slowly mixing dynamical systems, where an observable function has a polynomial decay of correlations against all $L^\infty$ test functions.
In the context of the large deviation theory~\cite{Ellis85}, the conjecture and Eq.~(\ref{large_deviations}) are suggestive;
the polynomial decay of large deviations is not usually taken into account even though $S_n(\phi)/n$ converges to the ensemble average;
furthermore, since the moment generating function (MGF) of the stable distribution for $0 < \alpha < 2$ can not be defined, one can not calculate the entropy function defined by the Legendre transform of the logarithm of MGF.
Improving the large deviation theory in the non-Gaussian regime $(1 < \alpha < 2)$ is a problem that remains to be solved.

Finally, we also remark that the power-laws in our model, as shown in Eqs.~(\ref{res_time_dis}) and (\ref{cor_fun}), arise from the functions $g(t)=t^B$.
However, it is not enough to explain the stagnant layer theory based on the Nekhoroshev theorem in nearly-integrable Hamiltonian systems~\cite{Aizawa89a,MorbidelliVergassola97,Nekhoroshev77}, where the probability density of the first passage time around tori obeys the log-Weibull expression,
\begin{equation*}
	f(m) \sim \frac{1}{m \left(\log m \right)^{1+d}} \qquad (m \gg 1),
\end{equation*}
where $d$ is a constant related to the degree of freedom.
From the viewpoint of infinite ergodic theory, we studied a one-dimensional map accompanied by the log-Weibull distribution~\cite{ShinkaiAizawa08}.
The results revealed that a logarithmic correction term, which is slowly varied, characterizes the extremely slow dynamics.
We should improve the function $g$ so that it affects the log-Weibull distribution around indifferent fixed points and has a uniform invariant measure.
This point will be reported elsewhere.

%%%%%%%%%%%%%%%%%%%%  ACKNOWLEDGMENT  %%%%%%%%%%%%%%%%%%%%%%%%%%%%%%%%%%%%%%%%
\begin{acknowledgments}
This work is a part of the outcome of research performed under a Waseda University Grant for Special Research Projects
(Project number: 2009A-872).
\end{acknowledgments}

\appendix
%%%%%%%%%%%%%%%%%%%%  APPENDIX A  %%%%%%%%%%%%%%%%%%%%%%%%%%%%%%%%%%%%%%%%
\section{The stable density $(1 < \alpha \le 2)$~\cite{Feller71}}\label{AppA}
The characteristic function of the stable distribution is given by
\begin{equation}\label{chara_stable}
	\tilde{v}(z; \alpha, \gamma, c) = \exp \left\{ -|cz|^\alpha e^{\pm i \pi \gamma / 2} \right\},
\end{equation}
where $0 < \alpha \le 2$ is called a stable exponent,
$\gamma$ is the skewness parameter~\footnote{The necessary and sufficient condition to be stable is
$|\gamma| \le \alpha$ for $0<\alpha<1$ and $|\gamma| \le 2-\alpha$ for $1<\alpha\le 2$.},
$c$ is a scale parameter and the $\pm$ sign depends on $z \gtrless 0$.
For $1 < \alpha < 2$ the series expansion of the stable density is given by
\begin{widetext}
\begin{equation*}
	v(x; \alpha, \gamma, c) = \left\{
	\begin{array}{cc}
		\ds -\frac{1}{\pi c} \, \sum_{k=1}^\infty \, \frac{(-x/c)^{k-1}}{k!} \,
		\ds \Gamma\left(1+\frac{k}{\alpha}\right)\,\sin \frac{\pi k}{2\alpha} (\gamma-\alpha) & (x >0),\\
		\ds \frac{1}{\pi c} \, \sum_{k=1}^\infty \, \frac{(x/c)^{k-1}}{k!} \,
		\ds \Gamma\left(1+\frac{k}{\alpha}\right)\,\sin \frac{\pi k}{2\alpha} (\gamma+\alpha) & (x <0).		
	\end{array}
	\right.
\end{equation*}
\end{widetext}
In this paper we denote the stable distribution by $V(x; \alpha, \gamma, c)$~\footnote{
Sometimes we write $v(x)$ and $V(x)$ without the parameters.}.
For $\alpha = 2$ the distribution corresponds to the Gauss distribution.

One of important properties of the stable distribution, which can be shown to be equivalent to its definition, is as follows:
Let $\{ \bm{X}_j \}$ be an arbitrary sequence of mutually independent random variables with a common distribution $V(x;\alpha,\gamma,c)$ and
$\bm{S}_n = \bm{X}_1 + \cdots + \bm{X}_n$.
Then,
\begin{equation}\label{stabledislaw}
	\mathrm{Pr} \, \left\{ \frac{\bm{S}_n}{n} n^{1-1/\alpha} \le x \right\} = V(x;\alpha,\gamma,c).
\end{equation}
Moreover, since the stable distribution trivially belongs to the domain of attraction of the distribution, the tail-sum varies regularly with exponent $-\alpha$:
\begin{equation}\label{property_stable}
	1 - V(x) + V(-x) \sim x^{-\alpha} h(x) \qquad (x \to \infty),
\end{equation}
where $h(x)$ is slowly varying at $\infty$.

%%%%%%%%%%%%%%%%%%%%  APPENDIX B  %%%%%%%%%%%%%%%%%%%%%%%%%%%%%%%%%%%%%%%%
\section{The algorithm to obtain the exponent $\alpha$ and the scale parameter $c$ of the stable density}\label{AppB}
Assuming Eq.~(\ref{stabledislaw}), the probability density of the random variable $\bm{S}_n/n$ is written as
\begin{equation*}
	p_n(x) = n^{1-1/\alpha} \, v\left(n^{1-1/\alpha} x\right),
\end{equation*}
and its characteristic function is given by
\begin{eqnarray*}
	\tilde{p}_n(z) = \int_{-\infty}^\infty e^{i z x} p_n(x) \, d x = u_n(z) + i w_n(z),
\end{eqnarray*}
where the functions $u_n$ and $w_n$ are defined as $\int \cos(zx) p_n(x) \, d x$ and $\int \sin(zx) p_n(x) \, d x$, respectively.
Using Eq.~(\ref{chara_stable}), for $z>0$ the characteristic function can be written as follows:
\begin{eqnarray*}
	\tilde{p}_n(z) & = & \exp \left\{ \ln r_n(z) + i \theta_n (z) \right\},\\
	\ln r_n(z) & = & - n^{1-\alpha} (c z)^\alpha \cos \frac{\pi \gamma}{2},\\
	\theta_n(z) & = & - n^{1-\alpha} (c z)^\alpha \sin \frac{\pi \gamma}{2},
\end{eqnarray*}
where the functions $r_n$ and $\theta_n$ are defined by
\begin{equation*}
	r_n(z) = \sqrt{u_n^2+w_n^2}, \quad \theta_n(z) = \arctan ( w_n / u_n).
\end{equation*}

If the function $\eta_n$ is defined by
\begin{equation*}
	\eta_n(z) = \sqrt{ (\ln r_n)^2 + \theta_n^2} = n^{1-\alpha} (c z)^\alpha,
\end{equation*}
we can use the graph of the log-log plot to evaluate it as follows:
\begin{equation}\label{log_eta}
	\ln \eta_n(z) = \alpha ( \ln z + \ln c) - (\alpha - 1) \ln n.
\end{equation}
Therefore, under the assumption that a numerical probability density is stable,
the exponent $\alpha$ and the scale parameter $c$ can be determined.


\begin{thebibliography}{35}%
\makeatletter
\providecommand \@ifxundefined [1]{%
 \@ifx{#1\undefined}
}%
\providecommand \@ifnum [1]{%
 \ifnum #1\expandafter \@firstoftwo
 \else \expandafter \@secondoftwo
 \fi
}%
\providecommand \@ifx [1]{%
 \ifx #1\expandafter \@firstoftwo
 \else \expandafter \@secondoftwo
 \fi
}%
\providecommand \natexlab [1]{#1}%
\providecommand \enquote  [1]{``#1''}%
\providecommand \bibnamefont  [1]{#1}%
\providecommand \bibfnamefont [1]{#1}%
\providecommand \citenamefont [1]{#1}%
\providecommand \href@noop [0]{\@secondoftwo}%
\providecommand \href [0]{\begingroup \@sanitize@url \@href}%
\providecommand \@href[1]{\@@startlink{#1}\@@href}%
\providecommand \@@href[1]{\endgroup#1\@@endlink}%
\providecommand \@sanitize@url [0]{\catcode `\\12\catcode `\$12\catcode
  `\&12\catcode `\#12\catcode `\^12\catcode `\_12\catcode `\%12\relax}%
\providecommand \@@startlink[1]{}%
\providecommand \@@endlink[0]{}%
\providecommand \url  [0]{\begingroup\@sanitize@url \@url }%
\providecommand \@url [1]{\endgroup\@href {#1}{\urlprefix }}%
\providecommand \urlprefix  [0]{URL }%
\providecommand \Eprint [0]{\href }%
\providecommand \doibase [0]{http://dx.doi.org/}%
\providecommand \selectlanguage [0]{\@gobble}%
\providecommand \bibinfo  [0]{\@secondoftwo}%
\providecommand \bibfield  [0]{\@secondoftwo}%
\providecommand \translation [1]{[#1]}%
\providecommand \BibitemOpen [0]{}%
\providecommand \bibitemStop [0]{}%
\providecommand \bibitemNoStop [0]{.\EOS\space}%
\providecommand \EOS [0]{\spacefactor3000\relax}%
\providecommand \BibitemShut  [1]{\csname bibitem#1\endcsname}%
\let\auto@bib@innerbib\@empty
%</preamble>
\bibitem [{\citenamefont {Aizawa}\ \emph {et~al.}(1989)\citenamefont {Aizawa},
  \citenamefont {Kikuchi}, \citenamefont {Harayama}, \citenamefont {Yamamoto},
  \citenamefont {Ota},\ and\ \citenamefont {Tanaka}}]{Aizawa89b}%
  \BibitemOpen
  \bibfield  {author} {\bibinfo {author} {\bibfnamefont {Y.}~\bibnamefont
  {Aizawa}}, \bibinfo {author} {\bibfnamefont {Y.}~\bibnamefont {Kikuchi}},
  \bibinfo {author} {\bibfnamefont {T.}~\bibnamefont {Harayama}}, \bibinfo
  {author} {\bibfnamefont {K.}~\bibnamefont {Yamamoto}}, \bibinfo {author}
  {\bibfnamefont {M.}~\bibnamefont {Ota}}, \ and\ \bibinfo {author}
  {\bibfnamefont {K.}~\bibnamefont {Tanaka}},\ }\href@noop {} {\bibfield
  {journal} {\bibinfo  {journal} {Prog. Theor. Phys. Suppl.}\ }\textbf
  {\bibinfo {volume} {98}},\ \bibinfo {pages} {36} (\bibinfo {year}
  {1989})}\BibitemShut {NoStop}%
\bibitem [{\citenamefont {Karney}(1983)}]{Karney83}%
  \BibitemOpen
  \bibfield  {author} {\bibinfo {author} {\bibfnamefont {C.~F.~F.}\
  \bibnamefont {Karney}},\ }\href@noop {} {\bibfield  {journal} {\bibinfo
  {journal} {Physica D}\ }\textbf {\bibinfo {volume} {8}},\ \bibinfo {pages}
  {360} (\bibinfo {year} {1983})}\BibitemShut {NoStop}%
\bibitem [{\citenamefont {Chirikov}\ and\ \citenamefont
  {Shepelyansky}(1984)}]{ChirikovShepelyansky84}%
  \BibitemOpen
  \bibfield  {author} {\bibinfo {author} {\bibfnamefont {B.~V.}\ \bibnamefont
  {Chirikov}}\ and\ \bibinfo {author} {\bibfnamefont {D.~L.}\ \bibnamefont
  {Shepelyansky}},\ }\href@noop {} {\bibfield  {journal} {\bibinfo  {journal}
  {Physica D}\ }\textbf {\bibinfo {volume} {13}},\ \bibinfo {pages} {395}
  (\bibinfo {year} {1984})}\BibitemShut {NoStop}%
\bibitem [{\citenamefont {Geisel}\ \emph {et~al.}(1987)\citenamefont {Geisel},
  \citenamefont {Zacherl},\ and\ \citenamefont
  {Radons}}]{GeiselZacherlRadons87}%
  \BibitemOpen
  \bibfield  {author} {\bibinfo {author} {\bibfnamefont {T.}~\bibnamefont
  {Geisel}}, \bibinfo {author} {\bibfnamefont {A.}~\bibnamefont {Zacherl}}, \
  and\ \bibinfo {author} {\bibfnamefont {G.}~\bibnamefont {Radons}},\
  }\href@noop {} {\bibfield  {journal} {\bibinfo  {journal} {Phys. Rev. Lett.}\
  }\textbf {\bibinfo {volume} {59}},\ \bibinfo {pages} {2503} (\bibinfo {year}
  {1987})}\BibitemShut {NoStop}%
\bibitem [{\citenamefont {Aizawa}(1989{\natexlab{a}})}]{Aizawa89a}%
  \BibitemOpen
  \bibfield  {author} {\bibinfo {author} {\bibfnamefont {Y.}~\bibnamefont
  {Aizawa}},\ }\href@noop {} {\bibfield  {journal} {\bibinfo  {journal} {Prog.
  Theor. Phys.}\ }\textbf {\bibinfo {volume} {81}},\ \bibinfo {pages} {249}
  (\bibinfo {year} {1989}{\natexlab{a}})}\BibitemShut {NoStop}%
\bibitem [{\citenamefont {Aizawa}\ \emph {et~al.}(2000)\citenamefont {Aizawa},
  \citenamefont {Sato},\ and\ \citenamefont {Ito}}]{AizawaSatoIto00}%
  \BibitemOpen
  \bibfield  {author} {\bibinfo {author} {\bibfnamefont {Y.}~\bibnamefont
  {Aizawa}}, \bibinfo {author} {\bibfnamefont {K.}~\bibnamefont {Sato}}, \ and\
  \bibinfo {author} {\bibfnamefont {K.}~\bibnamefont {Ito}},\ }\href@noop {}
  {\bibfield  {journal} {\bibinfo  {journal} {Prog. Theor. Phys.}\ }\textbf
  {\bibinfo {volume} {103}},\ \bibinfo {pages} {519} (\bibinfo {year}
  {2000})}\BibitemShut {NoStop}%
\bibitem [{\citenamefont {Aaronson}(1997)}]{Aaronson97}%
  \BibitemOpen
  \bibfield  {author} {\bibinfo {author} {\bibfnamefont {J.}~\bibnamefont
  {Aaronson}},\ }\href@noop {} {\emph {\bibinfo {title} {An Introduction to
  Infinite Ergodic Theory}}}\ (\bibinfo  {publisher} {American Mathematical
  Society},\ \bibinfo {year} {1997})\BibitemShut {NoStop}%
\bibitem [{\citenamefont {Aizawa}(2000)}]{Aizawa00}%
  \BibitemOpen
  \bibfield  {author} {\bibinfo {author} {\bibfnamefont {Y.}~\bibnamefont
  {Aizawa}},\ }\href@noop {} {\bibfield  {journal} {\bibinfo  {journal} {Chaos,
  Solitons and Fractals}\ }\textbf {\bibinfo {volume} {11}},\ \bibinfo {pages}
  {263} (\bibinfo {year} {2000})}\BibitemShut {NoStop}%
\bibitem [{\citenamefont {Shinkai}\ and\ \citenamefont
  {Aizawa}(2006)}]{ShinkaiAizawa06}%
  \BibitemOpen
  \bibfield  {author} {\bibinfo {author} {\bibfnamefont {S.}~\bibnamefont
  {Shinkai}}\ and\ \bibinfo {author} {\bibfnamefont {Y.}~\bibnamefont
  {Aizawa}},\ }\href@noop {} {\bibfield  {journal} {\bibinfo  {journal} {Prog.
  Theor. Phys.}\ }\textbf {\bibinfo {volume} {116}},\ \bibinfo {pages} {503}
  (\bibinfo {year} {2006})}\BibitemShut {NoStop}%
\bibitem [{\citenamefont {Akimoto}(2008)}]{Akimoto08}%
  \BibitemOpen
  \bibfield  {author} {\bibinfo {author} {\bibfnamefont {T.}~\bibnamefont
  {Akimoto}},\ }\href@noop {} {\bibfield  {journal} {\bibinfo  {journal} {J.
  Stat. Phys.}\ }\textbf {\bibinfo {volume} {132}},\ \bibinfo {pages} {171}
  (\bibinfo {year} {2008})}\BibitemShut {NoStop}%
\bibitem [{\citenamefont {Aizawa}(1989{\natexlab{b}})}]{Aizawa89c}%
  \BibitemOpen
  \bibfield  {author} {\bibinfo {author} {\bibfnamefont {Y.}~\bibnamefont
  {Aizawa}},\ }\href@noop {} {\bibfield  {journal} {\bibinfo  {journal} {Prog.
  Theor. Phys. Suppl.}\ }\textbf {\bibinfo {volume} {99}},\ \bibinfo {pages}
  {149} (\bibinfo {year} {1989}{\natexlab{b}})}\BibitemShut {NoStop}%
\bibitem [{\citenamefont {Kikuchi}\ and\ \citenamefont
  {Aizawa}(1990)}]{KikuchiAizawa90}%
  \BibitemOpen
  \bibfield  {author} {\bibinfo {author} {\bibfnamefont {Y.}~\bibnamefont
  {Kikuchi}}\ and\ \bibinfo {author} {\bibfnamefont {Y.}~\bibnamefont
  {Aizawa}},\ }\href@noop {} {\bibfield  {journal} {\bibinfo  {journal} {Prog.
  Theor. Phys.}\ }\textbf {\bibinfo {volume} {84}},\ \bibinfo {pages} {1014}
  (\bibinfo {year} {1990})}\BibitemShut {NoStop}%
\bibitem [{\citenamefont {Tanaka}\ and\ \citenamefont
  {Aizawa}(1993)}]{TanakaAizawa93}%
  \BibitemOpen
  \bibfield  {author} {\bibinfo {author} {\bibfnamefont {K.}~\bibnamefont
  {Tanaka}}\ and\ \bibinfo {author} {\bibfnamefont {Y.}~\bibnamefont
  {Aizawa}},\ }\href@noop {} {\bibfield  {journal} {\bibinfo  {journal} {Prog.
  Theor. Phys.}\ }\textbf {\bibinfo {volume} {90}},\ \bibinfo {pages} {547}
  (\bibinfo {year} {1993})}\BibitemShut {NoStop}%
\bibitem [{\citenamefont {Melbourne}(2009)}]{Melbourne09}%
  \BibitemOpen
  \bibfield  {author} {\bibinfo {author} {\bibfnamefont {I.}~\bibnamefont
  {Melbourne}},\ }\href@noop {} {\bibfield  {journal} {\bibinfo  {journal}
  {Proc. Am. Math. Soc.}\ }\textbf {\bibinfo {volume} {137}},\ \bibinfo {pages}
  {1735} (\bibinfo {year} {2009})}\BibitemShut {NoStop}%
\bibitem [{\citenamefont {Pollicott}\ and\ \citenamefont
  {Sharp}(2009)}]{PollicottSharp09}%
  \BibitemOpen
  \bibfield  {author} {\bibinfo {author} {\bibfnamefont {M.}~\bibnamefont
  {Pollicott}}\ and\ \bibinfo {author} {\bibfnamefont {R.}~\bibnamefont
  {Sharp}},\ }\href@noop {} {\bibfield  {journal} {\bibinfo  {journal}
  {Nonlinearity}\ }\textbf {\bibinfo {volume} {22}},\ \bibinfo {pages} {2079}
  (\bibinfo {year} {2009})}\BibitemShut {NoStop}%
\bibitem [{\citenamefont {Artuso}\ and\ \citenamefont
  {Manchein}(2009)}]{ArtusoManchein09}%
  \BibitemOpen
  \bibfield  {author} {\bibinfo {author} {\bibfnamefont {R.}~\bibnamefont
  {Artuso}}\ and\ \bibinfo {author} {\bibfnamefont {C.}~\bibnamefont
  {Manchein}},\ }\href@noop {} {\bibfield  {journal} {\bibinfo  {journal}
  {Phys. Rev. E}\ }\textbf {\bibinfo {volume} {80}},\ \bibinfo {pages} {036210}
  (\bibinfo {year} {2009})}\BibitemShut {NoStop}%
\bibitem [{\citenamefont {Pikovsky}(1991)}]{Pikovsky91}%
  \BibitemOpen
  \bibfield  {author} {\bibinfo {author} {\bibfnamefont {A.~S.}\ \bibnamefont
  {Pikovsky}},\ }\href@noop {} {\bibfield  {journal} {\bibinfo  {journal}
  {Phys. Rev. A}\ }\textbf {\bibinfo {volume} {43}},\ \bibinfo {pages} {3146}
  (\bibinfo {year} {1991})}\BibitemShut {NoStop}%
\bibitem [{\citenamefont {Miyaguchi}\ and\ \citenamefont
  {Aizawa}(2007)}]{MiyaguchiAizawa07}%
  \BibitemOpen
  \bibfield  {author} {\bibinfo {author} {\bibfnamefont {T.}~\bibnamefont
  {Miyaguchi}}\ and\ \bibinfo {author} {\bibfnamefont {Y.}~\bibnamefont
  {Aizawa}},\ }\href@noop {} {\bibfield  {journal} {\bibinfo  {journal} {Phys.
  Rev. E}\ }\textbf {\bibinfo {volume} {75}},\ \bibinfo {pages} {066201}
  (\bibinfo {year} {2007})}\BibitemShut {NoStop}%
\bibitem [{\citenamefont {Geisel}\ and\ \citenamefont
  {Thomae}(1984)}]{GeiselThomae84}%
  \BibitemOpen
  \bibfield  {author} {\bibinfo {author} {\bibfnamefont {T.}~\bibnamefont
  {Geisel}}\ and\ \bibinfo {author} {\bibfnamefont {S.}~\bibnamefont
  {Thomae}},\ }\href@noop {} {\bibfield  {journal} {\bibinfo  {journal} {Phys.
  Rev. Lett.}\ }\textbf {\bibinfo {volume} {52}},\ \bibinfo {pages} {1936}
  (\bibinfo {year} {1984})}\BibitemShut {NoStop}%
\bibitem [{\citenamefont {Klages}\ \emph {et~al.}(2008)\citenamefont {Klages},
  \citenamefont {Radons},\ and\ \citenamefont
  {Sokolov}}]{KlagesRadonsSokolv08}%
  \BibitemOpen
  \bibinfo {editor} {\bibfnamefont {R.}~\bibnamefont {Klages}}, \bibinfo
  {editor} {\bibfnamefont {G.}~\bibnamefont {Radons}}, \ and\ \bibinfo {editor}
  {\bibfnamefont {I.~M.}\ \bibnamefont {Sokolov}},\ eds.,\ \href@noop {} {\emph
  {\bibinfo {title} {Anomalous Transport: Foundations and Applications}}}\
  (\bibinfo  {publisher} {Wiley-VCH},\ \bibinfo {year} {2008})\BibitemShut
  {NoStop}%
\bibitem [{\citenamefont {Artuso}\ and\ \citenamefont
  {Cristandoro}(2004)}]{ArtusoCristadoro04}%
  \BibitemOpen
  \bibfield  {author} {\bibinfo {author} {\bibfnamefont {R.}~\bibnamefont
  {Artuso}}\ and\ \bibinfo {author} {\bibfnamefont {G.}~\bibnamefont
  {Cristandoro}},\ }\href@noop {} {\bibfield  {journal} {\bibinfo  {journal}
  {J. Phys. A}\ }\textbf {\bibinfo {volume} {37}},\ \bibinfo {pages} {85}
  (\bibinfo {year} {2004})}\BibitemShut {NoStop}%
\bibitem [{\citenamefont {Aizawa}(1984)}]{Aizawa84}%
  \BibitemOpen
  \bibfield  {author} {\bibinfo {author} {\bibfnamefont {Y.}~\bibnamefont
  {Aizawa}},\ }\href@noop {} {\bibfield  {journal} {\bibinfo  {journal} {Prog.
  Theor. Phys.}\ }\textbf {\bibinfo {volume} {72}},\ \bibinfo {pages} {659}
  (\bibinfo {year} {1984})}\BibitemShut {NoStop}%
\bibitem [{Note1()}]{Note1}%
  \BibitemOpen
  \bibinfo {note} {When $C(\tau )$ decays exponentially, the quantity
  corresponds to the relaxation time.}\BibitemShut {Stop}%
\bibitem [{\citenamefont {Feller}(1949)}]{Feller49}%
  \BibitemOpen
  \bibfield  {author} {\bibinfo {author} {\bibfnamefont {W.}~\bibnamefont
  {Feller}},\ }\href@noop {} {\bibfield  {journal} {\bibinfo  {journal} {Trans.
  Am. Math. Soc.}\ }\textbf {\bibinfo {volume} {67}},\ \bibinfo {pages} {98}
  (\bibinfo {year} {1949})}\BibitemShut {NoStop}%
\bibitem [{\citenamefont {Feller}(1971)}]{Feller71}%
  \BibitemOpen
  \bibfield  {author} {\bibinfo {author} {\bibfnamefont {W.}~\bibnamefont
  {Feller}},\ }\href@noop {} {\emph {\bibinfo {title} {An Introduction to
  Probability Theory and Its Applications}}},\ \bibinfo {edition} {2nd}\ ed.,\
  Vol.~\bibinfo {volume} {2}\ (\bibinfo  {publisher} {John Wiley \& Sons},\
  \bibinfo {year} {1971})\BibitemShut {NoStop}%
\bibitem [{\citenamefont {Arnold}\ and\ \citenamefont
  {Avez}(1968)}]{ArnoldAvez68}%
  \BibitemOpen
  \bibfield  {author} {\bibinfo {author} {\bibfnamefont {V.~I.}\ \bibnamefont
  {Arnold}}\ and\ \bibinfo {author} {\bibfnamefont {A.}~\bibnamefont {Avez}},\
  }\href@noop {} {\emph {\bibinfo {title} {Ergodic Problems of Classical
  Mechanics}}}\ (\bibinfo  {publisher} {W.A. Benjamin, New York},\ \bibinfo
  {year} {1968})\BibitemShut {NoStop}%
\bibitem [{\citenamefont {Godr\`eche}\ and\ \citenamefont
  {Luck}(2001)}]{GodrecheLuck01}%
  \BibitemOpen
  \bibfield  {author} {\bibinfo {author} {\bibfnamefont {C.}~\bibnamefont
  {Godr\`eche}}\ and\ \bibinfo {author} {\bibfnamefont {J.~M.}\ \bibnamefont
  {Luck}},\ }\href@noop {} {\bibfield  {journal} {\bibinfo  {journal} {J. Stat.
  Phys.}\ }\textbf {\bibinfo {volume} {104}},\ \bibinfo {pages} {489} (\bibinfo
  {year} {2001})}\BibitemShut {NoStop}%
\bibitem [{\citenamefont {Sinai}(2008)}]{Sinai08}%
  \BibitemOpen
  \bibfield  {author} {\bibinfo {author} {\bibfnamefont {Y.~G.}\ \bibnamefont
  {Sinai}},\ }\href@noop {} {\bibfield  {journal} {\bibinfo  {journal}
  {Nonlinearity}\ }\textbf {\bibinfo {volume} {21}},\ \bibinfo {pages} {T253}
  (\bibinfo {year} {2008})}\BibitemShut {NoStop}%
\bibitem [{\citenamefont {Gou\"ezel}(2008)}]{Gouezel08}%
  \BibitemOpen
  \bibfield  {author} {\bibinfo {author} {\bibfnamefont {S.}~\bibnamefont
  {Gou\"ezel}},\ }\href@noop {} {\enquote {\bibinfo {title} {Stable laws for
  the doubling map},}\ } (\bibinfo {year} {2008}),\ \bibinfo {note}
  {http://perso.univ-rennes1.fr/sebastien.gouezel/articles/DoublingStable.pdf
  (preprint)}\BibitemShut {NoStop}%
\bibitem [{\citenamefont {Ellis}(1985)}]{Ellis85}%
  \BibitemOpen
  \bibfield  {author} {\bibinfo {author} {\bibfnamefont {R.~S.}\ \bibnamefont
  {Ellis}},\ }\href@noop {} {\emph {\bibinfo {title} {Entropy, Large
  Deviations, and Statistical Mechanics}}}\ (\bibinfo  {publisher}
  {Springer-Verlag, New York},\ \bibinfo {year} {1985})\BibitemShut {NoStop}%
\bibitem [{\citenamefont {Morbidelli}\ and\ \citenamefont
  {Vergassola}(1997)}]{MorbidelliVergassola97}%
  \BibitemOpen
  \bibfield  {author} {\bibinfo {author} {\bibfnamefont {A.}~\bibnamefont
  {Morbidelli}}\ and\ \bibinfo {author} {\bibfnamefont {M.}~\bibnamefont
  {Vergassola}},\ }\href@noop {} {\bibfield  {journal} {\bibinfo  {journal} {J.
  Stat. Phys.}\ }\textbf {\bibinfo {volume} {89}},\ \bibinfo {pages} {549}
  (\bibinfo {year} {1997})}\BibitemShut {NoStop}%
\bibitem [{\citenamefont {Nekhoroshev}(1977)}]{Nekhoroshev77}%
  \BibitemOpen
  \bibfield  {author} {\bibinfo {author} {\bibfnamefont {N.~N.}\ \bibnamefont
  {Nekhoroshev}},\ }\href@noop {} {\bibfield  {journal} {\bibinfo  {journal}
  {Russ. Math. Surv.}\ }\textbf {\bibinfo {volume} {32}},\ \bibinfo {pages} {1}
  (\bibinfo {year} {1977})}\BibitemShut {NoStop}%
\bibitem [{\citenamefont {Shinkai}\ and\ \citenamefont
  {Aizawa}(2008)}]{ShinkaiAizawa08}%
  \BibitemOpen
  \bibfield  {author} {\bibinfo {author} {\bibfnamefont {S.}~\bibnamefont
  {Shinkai}}\ and\ \bibinfo {author} {\bibfnamefont {Y.}~\bibnamefont
  {Aizawa}},\ }in\ \href@noop {} {\emph {\bibinfo {booktitle} {Let's Face Chaos
  Through Nonlinear Dynamics}}}\ (\bibinfo  {publisher} {American Institute of
  Physics, New York},\ \bibinfo {year} {2008})\ pp.\ \bibinfo {pages}
  {219--22}\BibitemShut {NoStop}%
\bibitem [{Note2()}]{Note2}%
  \BibitemOpen
  \bibinfo {note} {The necessary and sufficient condition to be stable is
  $|\gamma | \le \alpha $ for $0<\alpha <1$ and $|\gamma | \le 2-\alpha $ for
  $1<\alpha \le 2$.}\BibitemShut {Stop}%
\bibitem [{Note3()}]{Note3}%
  \BibitemOpen
  \bibinfo {note} {Sometimes we write $v(x)$ and $V(x)$ without the
  parameters.}\BibitemShut {Stop}%
\end{thebibliography}
\end{document}